\documentclass{article}
\usepackage[a4paper,scale=0.8]{geometry}
\usepackage{amsmath}
\usepackage{amssymb}
\usepackage{amsthm}
\usepackage{bm}
\usepackage{arydshln}
\usepackage{url}
\usepackage[numbers,sort]{natbib}

\let\cite\citep

\newcommand{\vct}{\bm}

\newcommand{\bj}{{\vct{j}}}
\newcommand{\bk}{{\vct{k}}}

\newcommand{\bpi}{{\vct{\pi}}}
\newcommand{\bdelta}{{\vct{\delta}}}

\newcommand{\Nset}{{\{1,\dots,n\}}}
\newcommand{\Mset}{{\{1,\dots,m\}}}
\newcommand{\Vset}{{\{1,\dots,v\}}}

\newcommand{\R}{{\mathbb{R}}}
\newcommand{\Q}{{\mathbb{Q}}}
\newcommand{\Z}{{\mathbb{Z}}}
\newcommand{\A}{{\mathrm{A}}}
\newcommand{\B}{{\mathrm{B}}}
\newcommand{\X}{{\mathrm{X}}}
\newcommand{\K}{{\mathrm{K}}}
\newcommand{\trans}{{\mathrm{T}}}

\newcommand{\calW}{{\mathcal{W}}}

\newcommand{\CorP}{\mathrm{COR}^\square}
\newcommand{\CutP}{\mathrm{CUT}^\square}
\newcommand{\CutC}{\mathrm{CUT}}
\newcommand{\BellP}{\mathcal{B}^\square}
\newcommand{\sym}{{\mathcal{S}}}
\newcommand{\calG}{{\mathcal{G}}}
\newcommand{\calH}{{\mathcal{H}}}

\newcommand{\NP}{{\mathrm{NP}}}

\DeclareMathOperator{\conv}{conv}
\DeclareMathOperator{\cone}{cone}

\DeclareMathOperator{\aff}{aff}

\DeclareMathOperator{\nei}{N}
\DeclareMathOperator{\rank}{rank}
\DeclareMathOperator{\Aut}{Aut}
\DeclareMathOperator{\im}{im}

\makeatletter
\newcommand{\revddots}
  {\mathinner{\mkern1mu\raise\p@
   \vbox{\kern7\p@\hbox{.}}\mkern2mu
   \raise4\p@\hbox{.}\mkern2mu\raise7\p@\hbox{.}\mkern1mu}}
\makeatother

\newcommand{\sfperiod}{\spacefactor\the\sfcode`\.}

\theoremstyle{plain}
\newtheorem{theorem}{Theorem}[section]
\newtheorem{lemma}[theorem]{Lemma}
\newtheorem{corollary}[theorem]{Corollary}

\newtheorem{claim}{Claim}[section]

\theoremstyle{definition}
\newtheorem{definition}{Definition}[section]
\newtheorem{example}{Example}[section]

\theoremstyle{remark}
\newtheorem{remark}{Remark}[section]

\newcounter{enumroman}
\renewcommand{\theenumroman}{\roman{enumroman}}
\newenvironment{enumroman}
  {\list
     {(\theenumroman)}%
     {\usecounter{enumroman}\def\makelabel##1{\hss\llap{##1}}}}
  {\endlist}

\newenvironment{problem}
  {\begin{list}{}{\setlength{\itemsep}{0pt}\setlength{\parsep}{0pt}%
                  \setlength{\leftmargin}{2em}%
                  \setlength{\labelwidth}{0pt}%
                  \setlength{\itemindent}{-\leftmargin}%
                  \setlength{\listparindent}{\parindent}}}
  {\end{list}}

\title{Deriving Tight Bell Inequalities \\ for 2 Parties
  with Many 2-valued Observables \\ from Facets of Cut Polytopes}

\author{David Avis$^1$, Hiroshi Imai$^{2,3}$,
  Tsuyoshi Ito$^2$, and Yuuya Sasaki$^2$ \\[1ex]
  \normalsize
  \begin{tabular}{c@{}l}
    $^1$ & School of Computer Science, McGill University \\
         & 3480 University, Montreal, Quebec, Canada H3A 2A7. \\
    $^2$ & Department of Computer Science, University of Tokyo \\
         & 7-3-1 Hongo, Bunkyo-ku, Tokyo 113-0033, Japan. \\
    $^3$ & ERATO Quantum Computation and Information Project, Tokyo, Japan.
  \end{tabular}}

\date{April 18, 2004}

\begin{document}
\maketitle

\begin{abstract}

  Relatively few families of Bell inequalities have previously been
  identified.
  Some examples are
  the trivial, CHSH, $\mathrm{I}_{mm22}$, and CGLMP inequalities.
  This paper presents a large number of new families of tight Bell
  inequalities for the case of many observables.
  For example, 44,368,793 inequivalent tight Bell inequalities other
  than CHSH are obtained for the case of 2 parties each with 10
  2-valued observables.
  This is accomplished by first establishing a relationship between
  the Bell inequalities and the facets of the cut polytope, a well
  studied object in polyhedral combinatorics.
  We then prove a theorem allowing us to derive new facets of cut
  polytopes from facets of smaller polytopes by a process derived from
  Fourier-Motzkin elimination, which we call triangular elimination.
  These new facets in turn give new tight Bell inequalities.
  We give additional results for projections, liftings, and the
  complexity of membership testing for the associated Bell polytope.

\end{abstract}

\section{Introduction}

\paragraph*{Quantum nonlocality and Bell inequalities.}
Recently, it is strongly conjectured that the power of quantum
information theory over the classical one, such as unconditionally
secure secret communication, is based on a clever use of the quantum
nonlocality of states.
To explore what is possible in quantum information theory, it is
important to distinguish the quantum states which have nonlocality
from those which do not.

A quantum state has nonlocality if it produces a non-classical
correlation table as a result of the correlation experiment when each
party is given some set of observables.
This leads to the significance of the problem of testing whether a given
correlation table in a setting with $n$ parties each of which has $m$
$v$-valued observables is classical or not.

A linear inequality is called a \emph{Bell inequality} if it is
satisfied by all the classical correlation tables.
Bell inequalities are used to test a correlation table
because each of them is a necessary condition for a correlation table
to be classical.

\paragraph*{Complete facet list of Bell polytopes.}
Peres~\cite{Per:all99} showed that for $n,m,v\ge1$, all the classical
correlation tables in the $n$-party $m$-observable $v$-value setting
form a convex polytope, which we call the \emph{Bell polytope}
$\BellP(n,m,v)$.
An inequality is a Bell inequality if and only if it is valid for
$\BellP(n,m,v)$.

The membership test for $\BellP(n,m,v)$ corresponds to the problem of
testing whether a given correlation table is classical or not.
There is evidence suggesting that the membership test for
$\BellP(n,m,v)$ is computationally intractable~\cite{Pit-MP91}. Nevertheless
a knowledge of
valid inequalities for
$\BellP(n,m,v)$ 
allows us to demonstrate \emph{non-membership} efficiently: it is sufficient
to give a single violated Bell inequality
to show that the corresponding correlation table exhibits non-classical behavior.
Among valid inequalities, those which support facets of
$\BellP(n,m,v)$ are the most useful because all other valid
inequalities can be derived from them.
A Bell inequality is said to be \emph{tight} if and only if it supports a facet
of $\BellP(n,m,v)$.

Historically, Clauser, Horne, Shimony and Holt~\cite{ClaHorShiHol-PRL69}
introduced an
inequality valid for $\BellP(2,2,2)$, which is known as
the CHSH inequality.
Fine~\cite{Fin-PRL82} proved that the trivial inequalities and
inequalities equivalent to the CHSH inequality together form the
complete list of facets of $\BellP(2,2,2)$.
Because computing the facets of a high-dimensional polytope from its
vertices is a very difficult problem, the complete list of facets of
$\BellP(n,m,v)$ is currently known only for small $n$, $m$ and $v$,
namely for $(n,m,v)=(2,2,2)$ by Fine~\cite{Fin-PRL82},
$(2,3,2), (3,2,2)$ by Pitowsky and Svozil~\cite{PitSvo-PRA01},
and $(2,2,3)$ by Collins and Gisin~\cite{ColGis-JPA04}.
There is also the results of asymmetric settings of
observables~\cite{ColGis-JPA04,Sli-PLA03}.  These complete lists, in
general, consist of a large number of inequalities.  These lists are
symmetric with respect to the exchange of parties, observables and
values (see e.g.\
\cite{WerWol-PRA01,Mas-QIC03,Sli-PLA03,ColGis-JPA04}).
\'{S}liwa~\cite{Sli-PLA03} and Collins and Gisin~\cite{ColGis-JPA04}
independently classified the facets of
$\BellP(n,m,v)$ according to these symmetries for small $n$, $m$ and
$v$, and showed, for instance, that $\BellP(2,3,2)$ have only $3$
inequivalent facets, the trivial, the CHSH and $I_{3322}$.

\paragraph*{Other known Bell inequalities.}
The difficulty of finding a complete facet characterization for
$\BellP(n,m,v)$ opens two directions of study.
One is to find some, instead of all, of the facets of $\BellP(n,m,v)$.
In this direction, Collins and Gisin~\cite{ColGis-JPA04} show a
family $\mathrm{I}_{mm22}$ of inequalities valid for $\BellP(2,m,2)$
for general $m$, which are confirmed to support a facet for $m\le7$.
Masanes~\cite{Mas-QIC03} shows that the CGLMP
inequalities~\cite{ColGisLinMasPop-PRL02} valid for $\BellP(2,2,v)$
actually support facets of $\BellP(2,2,v)$.

The other direction is to study the complete list of facets of an
affinely projected image of $\BellP(n,m,v)$ and lift them to a valid
inequalities for $\BellP(n,m,v)$.
Werner and Wolf~\cite{WerWol-PRA01} consider the polytope formed by
the full correlation functions,
which is viewed as a kind of projection (see
Section~\ref{sbsect:proj-correlation-function}).

For other results about Bell inequalities, see a
survey paper by Werner and Wolf~\cite{WerWol-QIC01}.

\paragraph*{Correlation polytopes and cut polytopes.}
In \cite{Pit-MP91}, Pitowsky introduced correlation polytopes as the
set of possible joint correlations of events in a probabilistic space,
and showed the equivalence of $\BellP(2,2,2)$ to the correlation
polytope $\CorP(\K_{2,2})$ of a complete bipartite graph $\K_{2,2}$.
Cut polytopes, which have essentially the same structure as the
correlation polytopes, were independently introduced in combinatorial
optimization and have been extensively studied.
Many results are known for cut polytopes, including their
relation to correlation polytopes, which suggests we consider Bell
polytopes in the context of cut polytopes.
The book by Deza and Laurent~\cite{DezLau:cut97} is a comprehensive
study of these polyhedra, and their applications.
Ziegler~\cite{Zie:lectures98} is a good source for basic definitions and
results on
convex polyhedra.

\paragraph*{Results in the case $n=v=2$.}
In the direction of finding a partial list of facets of $\BellP(n,m,v)$,
we consider the case of $n=v=2$.
Restricting the setting to $n=v=2$ has the advantage that
$\BellP(2,m,2)$ is affinely isomorphic to the cut polytope
$\CutP(\K_{1,m,m})$ of a complete tripartite graph $\K_{1,m,m}$,
which we prove in Section~\ref{sect:bell-cut}.
We give an operation named triangular elimination which transforms a
facet of a cut polytope of one graph to a facet of a cut polytope of a
larger graph.
This operation transforms a facet of a cut polytope of the complete graph
to a facet of $\CutP(\K_{1,m,m})$, which can be transformed to a facet
of $\BellP(2,m,2)$ through an affine isomorphism.
By using this operation and the list of known facets of
$\CutP_9$~\cite{Smapo}, we have 44,368,793 different non-CHSH facets
of $\BellP(2,10,2)$ of which no two inequalities are equivalent up to
the exchange of parties, observables or values.
In addition, two additional results are proved:
(1) A facet-supporting inequality for $\BellP(2,m,2)$ also supports a
    facet of $\BellP(2,m',2)$ for any $m'>m$.
(2) The membership test of $\BellP(2,m,2)$ is $\NP$-complete, which
    strengthens the unlikeliness of the complete facet
    characterization.

\paragraph*{Results on projections and liftings of $\BellP(n,m,v)$.}
In the direction of studying an affinely projected image of
$\BellP(n,m,v)$, we consider projections other than that to the
polytope formed by the full correlation functions.
We prove that a facet-supporting inequality for
$\BellP(n,m,v)$ never supports a facet of $\BellP(n',m,v)$ for any
$n'>n$, in contrast to the case of $\BellP(2,m',2)$.

\paragraph*{Result on dimension of $\BellP(n,m,v)$.}
In addition, we identify the dimension of $\BellP(n,m,v)$
in Section~\ref{sect:definition} by proving that the only linear
equations valid for $\BellP(n,m,v)$ are the normalization condition
and the no-signaling condition.

\section{Definition and dimension of Bell polytope}
  \label{sect:definition}

Peres~\cite{Per:all99} shows that for $n,m,v\ge1$, all the classical
correlation tables in a setting with $n$ parties each of which has $m$
$v$-valued observables form a polytope defined as follows.

\begin{definition}[Bell polytope of $(n,m,v)$-setting]
    \label{defn:bell-polytope}
  The \emph{Bell polytope} of $(n,m,v)$-setting is defined as the
  convex hull of $v^{nm}$ points $\vct{\beta}(\vct{c})$:
  \[
    \BellP(n,m,v) = \conv\{\vct{\beta}(\vct{c}) \in \R^{(mv)^n} \mid
                           \vct{c} \in \Vset^{n\times m}\},
  \]
  where
  \[
    \beta(\vct{c})_{(j_1,k_1),\dots,(j_n,k_n)} = \begin{cases}
      1 & \text{if $c_{ij_i}=k_i$ for all $1\le i\le n$,} \\
      0 & \text{otherwise.}
    \end{cases}
  \]
  Here the $(mv)^n$ coordinates of vectors in $\R^{(mv)^n}$ are
  indexed by
  $(j_1,k_1),\dots,(j_n,k_n)\in(\{1,\dots,m\}\times\{1,\dots,v\})^n$.
\end{definition}

$\BellP(n,m,v)$ is not of full dimension.
It is straightforward to show that any point $\vct{q}\in\BellP(n,m,v)$
satisfies the following linear equations.
\begin{itemize}
  \item
    \emph{Normalization condition}:
    For each $\bj \in \Mset^{n}$,
    \begin{equation}
      \sum_{\bk \in \Vset^{n}} q_{(j_{1},k_{1}), \ldots, (j_{n},k_{n})} = 1.
      \label{eq:normalization-condition}
    \end{equation}
  \item
    \emph{No-signaling condition}:
    For each $i^*\in \Nset$,
    $j_1,\dots,j_{i^{*}-1},j_{i^{*}+1},\dots,j_n,j,j'\in \Mset$, \break
    $k_1,\dots,k_{i^{*}-1},k_{i^{*}+1},\dots,k_n\in \Vset$,
    \begin{equation}
       \sum_{k=1}^v q_{(j_1,k_1),\dots,(j,k),\dots,(j_n,k_n)}
      =\sum_{k=1}^v q_{(j_1,k_1),\dots,(j',k),\dots,(j_n,k_n)}.
      \label{eq:no-signaling-condition}
    \end{equation}
\end{itemize}

The following theorem states that $\BellP(n,m,v)$ is of full
dimensional in the affine subspace defined by these equations.

\begin{theorem} \label{thrm:bell-dim}
  For $n,m,v\ge1$,
  \[
    \dim\BellP(n,m,v)=\bigl(m(v-1)+1\bigr)^n-1.
  \]
\end{theorem}

\begin{proof}
  Equations~\eqref{eq:normalization-condition} and
  \eqref{eq:no-signaling-condition} together define
  an $\bigl(\bigl(m(v-1)+1\bigr)^n-1\bigr)$-dimensional
  affine subspace of $\R^{(mv)^n}$ in which $\BellP(n,m,v)$ lies.
  This means $\dim\BellP(n,m,v)\le\bigl(m(v-1)+1\bigr)^n-1$.

  On the other hand, we can find $\bigl(m(v-1)+1\bigr)^n$ affinely independent
  points in $\BellP(n,m,v)$ as follows.
  Let $L=(\{1,\dots,m\}\times\{1,\dots,v-1\})\cup\{{*}\}$.
  Note that $\lvert L\rvert=m(v-1)+1$.
  For any $l_1,\dots,l_n\in L$,
  define $\vct{c}^{(l_1,\dots,l_n)}\in\{1,\dots,v\}^{n\times m}$ by
  \[
    c_{ij}^{(l_1,\dots,l_n)}=\begin{cases}
      k_i & \text{if $l_i=(j,k_i)$,} \\
       v  & \text{if $l_i={*}$ or $l_i=(j',k_i)$ for some $j'\neq j$.}
    \end{cases}
  \]
  Let $\vct{q}^{(l_1,\dots,l_n)}=\vct{\beta}(\vct{c}^{(l_1,\dots,l_n)})$.
  By definition of $\BellP(n,m,v)$, all of the $\bigl(m(v-1)+1\bigr)^n$ points
  $\vct{q}^{(l_1,\dots,l_n)}$ belong to $\BellP(n,m,v)$.
  It is straightforward that these $\bigl(m(v-1)+1\bigr)^n$ points
  are affinely independent.
  This means $\dim\BellP(n,m,v)\ge\bigl(m(v-1)+1\bigr)^n-1$,
  hence $\dim\BellP(n,m,v)=\bigl(m(v-1)+1\bigr)^n-1$.
\end{proof}

\section{Affine isomorphism between $\BellP(2,m,2)$ and $\CutP(K_{1,m,m})$}
  \label{sect:bell-cut}

In this section, we restrict our focus to the case of $n=v=2$.
In this setting we can give an affine isomorphism between
the Bell polytope $\BellP(2,m,2)$ and the cut polytope $\CutP(K_{1,m,m})$.
Cut polytopes have been extensively studied in combinatorial geometry
both theoretically and computationally,
and we can use these results to study Bell polytopes $\BellP(2,m,2)$.

We begin by giving definitions of the correlation polytopes and cut
polytopes.

\begin{definition}[Correlation polytope~%
    {\cite[Section~5.1]{DezLau:cut97}}]
  Let $G=(V,E)$ be a graph with vertex set $V$ and edge set $E$ where
  an edge connecting vertices $u$ and $v$ is denoted by $uv$.
  Consider an $\R$-vector space $\R^{V\cup E}$,
  that is, a $(\lvert V\rvert+\lvert E\rvert)$-dimensional vector space
  over $\R$ whose coordinates are labeled by $V\cup E$.

  We define the \emph{correlation polytope} of the graph $G$ as the
  convex hull of the points $\vct{\pi}(I)$:
  \[
    \CorP(G) = \conv\{\bpi(I) \mid I \subseteq V\},
  \]
  where
  \[
    \text{for $v\in V$, }
    \pi_{v}(I) = \begin{cases}
      1 & \text{if $v \in I$,} \\
      0 & \text{otherwise,}
    \end{cases} \qquad
    \text{and for $uv\in E$, }
    \pi_{uv}(I) = \begin{cases}
      1 & \text{if $\{u,v\} \subseteq I$,} \\
      0 & \text{otherwise.}
    \end{cases}
  \]
\end{definition}

\begin{definition}[Cut polytope~{\cite[Section~4.1]{DezLau:cut97}}]
    \label{defn:cut-polytope}
  Let $G=(V,E)$ be a graph.
  Consider an $\R$-vector space $\R^E$.
  The \emph{cut polytope} of the graph $G$ is defined as the convex
  hull of the points $\vct{\delta}(I)$:
  \[
    \CutP(G) = \conv\{\bdelta(I) \mid I \subseteq V\},
  \]
  where for $uv\in E$,
  \[
    \delta_{uv}(I) = \begin{cases}
      1 & \text{if exactly one of $u$ and $v$ is in $I$,} \\
      0 & \text{otherwise.}
    \end{cases}
  \]
  For simplicity, we denote the cut polytope $\CutP(\K_n)$
  of a complete graph $\K_n$ by $\CutP_n$.
\end{definition}

The cut polytope is highly symmetric.
For example, it looks the ``same'' at each vertex.
This notion is formalized by the switching operation.

\begin{definition}[Switching]
  \label{defn:switching}
  For a vector $\vct{x}\in\R^E$ and
  a set $S\subseteq V$,
  the \emph{switching} of $\vct{x}$ by $S$ is the vector $\vct{x}'\in\R^E$
  defined by
  \[
    x'_{uv}=\begin{cases}
      -x_{uv} & (\delta_{uv}(S)=1) \\
       x_{uv} & (\delta_{uv}(S)=0),
    \end{cases}
  \]
  and denoted by $\vct{x}^S$.

  For a vector $\vct{a}\in\R^E$, a scalar $a_0\in\R$
  and a set $S\subseteq V$,
  the \emph{switching} of the inequality $\vct{a}^\trans\vct{x}\le a_0$ by $S$
  is the inequality $(\vct{a}^S)^\trans\vct{x}\le a_0-\vct{a}^\trans\delta(S)$.
  This switching is valid for $\CutP(G)$
  if and only if the original inequality $\vct{a}^\trans\vct{x}\le a_0$
  is valid for $\CutP(G)$.   
  Similarly, the switching supports a facet of $\CutP(G)$
  if and only if the original inequality supports a facet of $\CutP(G)$.

  For a facet $f$ of $\CutP(G)$ supported
  by the inequality $\vct{a}^\trans\vct{x}\le a_0$,
  the \emph{switching} of $f$ by $S$ is the facet of $\CutP(G)$
  supported by the switching of $\vct{a}^\trans\vct{x}\le a_0$ by $S$,
  and denoted by $\gamma(S)\cdot f$.
\end{definition}

Pitowsky~\cite[pp.~27--29]{Pit:prob89} shows that the Bell polytope
$\BellP(2,2,2)$ is affinely isomorphic to the correlation polytope
$\CorP(\K_{2,2})$ of a complete bipartite graph $\K_{2,2}$.
The same applies to $\BellP(2,m,2)$ and $\CorP(\K_{m,m})$ for general
$m$ as the next theorem states.

\begin{theorem} \label{thrm:bell-cor}
  $\BellP(2,m,2)$ is affinely isomorphic to
  $\CorP(\K_{m,m})$, where $\K_{m,m}$ is the complete bipartite
  graph with $m$ left vertices and $m$ right vertices.
  The isomorphism maps every $\vct{q}\in\BellP(2,m,2)$ to
  $\vct{p}\in\CorP(\K_{m,m})$ given by:
  \begin{align}
    p_{j_{1},j_{2}+m}&=q_{(j_{1},2),(j_{2},2)} \nonumber \\
    p_{j_{1}}&=q_{ (j_{1},2),(j_{2},2)}+q_{(j_{1},2),(j_{2},1)}
      \label{eq:bell-cor-2} \\
    p_{j_{2}+m}&=q_{ (j_{1},2),(j_{2},2)}+q_{(j_{1},1),(j_{2},2)},
      \label{eq:bell-cor-3}
  \end{align}
  where $1\le j_1,j_2\le m$.
  Note that the value of the right hand side of
  equation~\eqref{eq:bell-cor-2} does not depend on $j_2$,
  and the value of the right hand side of
  equation~\eqref{eq:bell-cor-3} does not depend on $j_1$.
\end{theorem}

\begin{definition}[Suspension graph]
  Let $G=(V,E)$ be a graph with $n$ vertices.
  Then the \emph{suspension graph} $\nabla G$ of $G$ is
  the graph $G'=(V\cup\{{0}\},E\cup\{ 0v \mid v\in V\})$
  with $n+1$ vertices obtained from $G$
  by adding one new vertex
  and connecting it to each of $n$ existing vertices.
\end{definition}

\begin{theorem}[Affine isomorphism of $\CorP(G)$ to $\CutP(\nabla G)$
    {\cite[Section~5.2]{DezLau:cut97}}]
  \label{thrm:cor-cut}
  For any graph $G=(V,E)$, the correlation polytope $\CorP(G)$ of $G$
  is affinely isomorphic to the cut polytope $\CutP(\nabla G)$
  of the suspension graph of $G$.
  The isomorphism maps every $\vct{p}\in\CorP(G)$
  to $\vct{x}\in\CutP(\nabla G)$ given by:
  \begin{alignat*}{2}
    x_{0v}&=p_v & \quad &(v\in V), \\
    x_{uv}&=p_u+p_v-2p_{uv} & \quad &(uv\in E).
  \end{alignat*}
\end{theorem}

Combining Theorems~\ref{thrm:bell-cor} and \ref{thrm:cor-cut}
give the following result immediately.
Note that $\nabla\K_{m,m}=\K_{1,m,m}$.

\begin{theorem} \label{thrm:bell-cut}
  $\BellP(2,m,2)$ is affinely isomorphic to $\CutP(\K_{1,m,m})$,
  where $\K_{1,m,m}$ is the complete tripartite graph
  with one partition with one vertex and two partitions
  with $m$ vertices each.
  The isomorphism maps every $\vct{q}\in\BellP(2,m,2)$
  to $\vct{x}\in\CutP(\K_{1,m,m})$ given by:
  \begin{align}
    x_{0,j_1}&=q_{(j_1,2),(j_2,2)}+q_{(j_1,2),(j_2,1)},
      \label{eq:bell-cut-1} \\
    x_{0,j_2+m}&=q_{(j_1,2),(j_2,2)}+q_{(j_1,1),(j_2,2)},
      \label{eq:bell-cut-2} \\
    x_{j_1,j_2+m}&=q_{(j_1,2),(j_2,1)}+q_{(j_1,1),(j_2,2)}, \nonumber
  \end{align}
  where $1\le j_1,j_2\le m$.
  Note that the value of the right hand side of equation \eqref{eq:bell-cut-1}
  does not depend on $j_2$,
  and the value of the right hand side of equation \eqref{eq:bell-cut-2}
  does not depend on $j_1$.
\end{theorem}

Thus, the study of the Bell polytope $\BellP(2,m,2)$
is equivalent to that of the cut polytope $\CutP(\K_{1,m,m})$.
Both correlation polytopes and cut polytopes are of full dimension.
This means each of their facets has a unique representation by
a supporting linear inequality.

The cut polytopes of complete graphs have been extensively studied and
large classes of their facets are known.
In addition, conjectured complete lists of facets of $\CutP_n$ is
known for $n\le 9$~\cite{Smapo}, of which the lists for $n\le7$ are
known to be complete.
It is also known that the problem of testing membership in
$\CutP_n$ is known $\NP$-complete~\cite{AviDez-Net91}, so a
complete facet characterization for general $n$ is unlikely.

\section{Triangular elimination}

In this section, we establish a method called \emph{triangular
elimination} to construct facets of $\CutP(\K_{1,m,m})$ from facets of
$\CutP_n$ where
$m\ge\frac12\lfloor\frac{n-2}{2}\rfloor\lfloor\frac{n-4}{2}\rfloor$.

When $2m+1>n$, the zero-lifting theorem~\cite{Des-ORL90} guarantees
that any facet-supporting inequality for $\CutP_n$ supports a facet of
$\CutP_{2m+1}$.
Since $\CutP(\K_{1,m,m})$ is a projected image of $\CutP_{2m+1}$,
Fourier-Motzkin elimination~\cite[Lecture~1]{Zie:lectures98} can be used to convert
facet-supporting inequalities for $\CutP_{2m+1}$ to valid inequalities
for $\CutP(\K_{1,m,m})$.
The problem is that Fourier-Motzkin elimination does not always
produce facet-supporting inequalities.
The triangular elimination procedure we introduce here is a sufficient condition
for Fourier-Motzkin elimination to produce facet-supporting
inequalities.

Theorem~\ref{thrm:kn-k1mm} is the main theorem in this section which
guarantees triangular elimination always produce a facet of
$\CutP(\K_{1,m,m})$ from a non-triangle facet of $\CutP_n$.
It relies strongly on Theorem~\ref{thrm:facet-preserving}, which is
likely to be of independent interest to researchers interested in
polyhedral theory.

\subsection{Definition of triangular elimination}

Consider a graph $G=(V,E)$.
Let $u,u'\in V$, $uu'\in E$, and $A\subseteq\nei_G(u)\cap\nei_G(u')$,
where $\nei_G(v)$ is the set of vertices adjacent to $v\in V$ in $G$.
We create a graph $G^+=(V^+,E^+)$ from $G$
by removing the edge $uu'$ and instead adding a new vertex $v$ adjacent to
each vertex of $\{u,u'\}\cup A$:
\[
  V^+=V\cup\{v\},\quad
  E^+=(E\setminus\{uu'\})\cup\{uv,u'v\}\cup\{vw\mid w\in A\}.
\]

\begin{definition}[Detour extension]
  We say such $G^+$ is a \emph{detour extension} of $G$
  with removed edge $uu'$, added vertex $v$ and adjacent vertex set $A$.
\end{definition}

We would like to construct a facet of $\CutP(G^+)$ from a facet of $\CutP(G)$.
Let $G'=(V^+,E^+\cup\{uu'\})$.
Note that we can always construct a facet of $\CutP(G')$
from a facet $f$ of $\CutP(G)$
by the following zero-lifting theorem.

\begin{theorem}[Zero-lifting theorem~\cite{Des-ORL90}]
  Let $G=(V,E)$ be a graph, $u\in V$ and $A\subseteq\nei_G(u)$.
  We create a graph $G'=(V',E')$ from $G$
  by adding a new vertex $v$ adjacent to each vertex of $\{u\}\cup A$:
  \[
    V'=V\cup\{v\},\quad
    E'=E\cup\{uv\}\cup\{vw\mid w\in A\}.
  \]
  For an inequality $\vct{a}^\trans\vct{x}\le a_0$ in $\R^E$, define
  its \emph{zero-lifting} $\bar{\vct{a}}^\trans\vct{x}\le a_0$ by
  \[
    \bar{a}_{ww'}=\begin{cases}
      a_{ww'} & \text{if $ww'\in E$,} \\
        0     & \text{otherwise.}
    \end{cases}
  \]
  If the inequality $\vct{a}^\trans\vct{x}\le a_0$ supports a facet of
  $\CutP(G)$, then its zero-lifting supports a facet of $\CutP(G')$.
\end{theorem}

Let $\vct{a}^\trans\vct{x}\le a_0$ be the inequality supporting $f$.
If $a_{uu'}=0$, the zero-lifting of $\vct{a}^\trans\vct{x}\le a_0$ is
not only an inequality in $\R^{E'}$ but also an inequality in
$\R^{E^+}$.
In this case, the zero-lifting theorem guarantees it supports a facet
of $\CutP(G^+)$.
How can we extend this construction to the case of $a_{uu'}\neq0$?
The answer is to \emph{eliminate} the term $a_{uu'}x_{uu'}$
from the inequality $\vct{a}^\trans\vct{x}\le a_0$
by adding an appropriate valid inequality to it.
We consider the simplest case of adding a triangle inequality
$-a_{uu'}x_{uu'}+a_{uu'}x_{uv}-\lvert a_{uu'}\rvert x_{u'v}\le0$.
We call this operation the \emph{triangular elimination}.

The formal definition is as follows.

\begin{definition}[Triangular elimination]
  Let $G=(V,E)$ be a graph,
  and let $G^+=(V^+,E^+)$ be the detour extension of $G$
  with removed edge $uu'\in E$, added vertex $v$ and adjacent vertex set $A$.
  Let $\vct{a}^\trans\vct{x}\le a_0$ be an inequality in $\R^E$.
  The \emph{triangular elimination} of the inequality
  $\vct{a}^\trans\vct{x}\le a_0$ is the inequality in $\R^{E^+}$ defined by
  \begin{equation}
    \vct{a}^\trans\vct{x}
      -a_{uu'}x_{uu'}+a_{uu'}x_{uv}-\lvert a_{uu'}\rvert x_{u'v}\le a_0.
    \label{eq:triangular-elimination}
  \end{equation}
  Note that \eqref{eq:triangular-elimination} is indeed an inequality
  in $\R^{E^+}$ because it does not have the term of $x_{uu'}$.
\end{definition}

\subsection{Properties of inequality produced by triangular elimination}

The cut cone $\CutC(G)$ of a graph $G$ is a polyhedral cone closely
related to the cut polytope $\CutP(G)$.
The set of facets of $\CutC(G)$ consists of the facets of $\CutP(G)$
containing the coordinate origin.
In this subsection, we begin by considering the case of cut cones
because they are easier for theoretical handling, and then make use of
the switching operation to extend the results to cut polytopes.

\begin{definition}[Cut cone~{\cite[Section~4.1]{DezLau:cut97}}]
  \label{defn:cut-cone}
  Let $G=(V,E)$ be a graph.
  Consider an $\R$-vector space $\R^E$.
  The \emph{cut cone} of the graph $G$ is defined as the conic hull
  of the vectors $\vct{\delta}(I)$:
  \begin{align*}
    \CutC(G) &= \cone\{\bdelta(I) \mid I \subseteq V\} \\
             &= \Bigl\{\sum_{I\subseteq V}\lambda_I\vct{\delta}(I) \Bigm|
                       \lambda_I\le0\;(\forall I\subseteq V)\Bigr\},
  \end{align*}
  where $\vct{\delta}(I)$ is the same as in
  Definition~\ref{defn:cut-polytope}.
  Like the cut polytope, we denote the cut cone $\CutC(\K_n)$ of a
  complete graph by $\CutC_n$.
\end{definition}

First we show that the inequality produced by triangular elimination
from a valid inequality for $\CutC(G)$ is valid for $\CutC(G^+)$.

\begin{theorem} \label{thrm:valid}
  Let $G=(V,E)$ be a graph,
  and let $G^+=(V^+,E^+)$ be the detour extension of $G$
  with removed edge $uu'\in E$, added vertex $v$ and adjacent vertex set $A$.
  Let $\vct{a}^\trans\vct{x}\le a_0$ be an inequality in $\R^E$,
  and $\vct{b}^\trans\vct{x}\le a_0$ be its triangular elimination.
  Then, the following two conditions are equivalent.
  \begin{enumroman}
    \item The inequality $\vct{a}^\trans\vct{x}\le a_0$ is valid
      for $\CutC(G)$.
    \item The inequality $\vct{b}^\trans\vct{x}\le a_0$ is valid
      for $\CutC(G^+)$.
  \end{enumroman}
\end{theorem}

\begin{proof}
  \noindent $(\text{(i)}\implies\text{(ii)})$ \quad
  Let $G'=(V^+,E^+\cup\{uu'\})$.
  The inequality $\vct{b}^\trans\vct{x}\le a_0$
  is the sum of two inequalities
  \[
    \vct{a}^\trans\vct{x}\le a_0, \quad
    -a_{uu'}x_{uu'}+a_{uu'}x_{uv}-\lvert a_{uu'}\rvert x_{u'v}\le0,
  \]
  both of which are valid for $\CutC(G')$.
  This means the inequality $\vct{b}^\trans\vct{x}\le a_0$ is
  also valid for $\CutC(G')$, hence valid for $\CutC(G^+)$.

  \noindent $(\text{(ii)}\implies\text{(i)})$ \quad
  $\vct{a}^\trans\vct{x}$ is obtained from $\vct{b}^\trans\vct{x}$
  by collapsing two vertices $u'$ and $v$.
  This implies $\text{(ii)}\implies\text{(i)}$.
\end{proof}

\begin{remark} \label{remk:valid}
  As far as the validity of inequalities is concerned,
  we do not need the condition $A\subseteq\nei_G(u)\cap\nei_G(u')$
  for detour extension.
  However, this condition is needed to preserve the facet-supporting
  property of the inequalities, which we consider next.
\end{remark}

\begin{theorem} \label{thrm:facet-preserving}
  Let $G=(V,E)$ be a graph,
  and let $G^+=(V^+,E^+)$ be the detour extension of $G$
  with removed edge $uu'\in E$, added vertex $v$ and adjacent vertex set $A$.
  Let $\vct{a}^\trans\vct{x}\le0$ be an inequality
  supporting a facet of $\CutC(G)$.
  If there exists an edge $e\in E\setminus(\{uu'\}\cup\{uw,u'w\mid w\in A\})$
  such that $a_e\neq0$, then the triangular elimination of
  $\vct{a}^\trans\vct{x}\le0$ supports a facet of $\CutC(G^+)$.
\end{theorem}

Note that Lemma~\ref{lemm:projection} states essentially the same
thing as Lemma~26.5.2~(ii) of \cite{DezLau:cut97}.

\begin{lemma}
  \label{lemm:projection}
  Let $\R^E$ be a vector space of a finite dimension, and let $D\subseteq E$.
  Let $\pi\colon\R^E\to\R^D$ be the orthogonal projection.
  Let $P$ be a full-dimensional polyhedron in $\R^E$.
  Let $f$ be a facet of $P$ supported by an inequality
  $\vct{a}^\trans\vct{x}\le a_0$.
  If there exists $e\in E\setminus D$ such that $a_e\neq0$,
  then the projected image $\pi(f)$ is of full dimension in $\R^D$.
\end{lemma}

\begin{proof}
  Let $H$ be the hyperplane defined by $\vct{a}^\trans\vct{x}=a_0$,
  so that $f=P\cap H$.
  We prove $\pi(H)=\R^D$.
  For any $\vct{y}\in\R^D$, define $\vct{x}\in\R^E$ as follows.
  Let $x_{e'}=y_{e'}$ for each $e'\in D$.
  Set any values to $x_{e'}$ for $e'\in E\setminus(D\cup\{e\})$.
  Set $x_e$ by
  \[
    x_e=a_0-\sum_{e'\in E\setminus\{e\}}a_{e'}x_{e'}.
  \]
  This $\vct{x}$ is on $H$ and satisfies
  $\pi(\vct{x})=\vct{y}$.
  This means $\vct{y}\in\pi(H)$, which means $\pi(H)=\R^D$.

  Therefore, $\pi(f)=\pi(P)\cap\pi(H)=\pi(P)$
  is of full dimension in $\R^D$.
\end{proof}

\begin{proof}[Proof of Theorem~\ref{thrm:facet-preserving}]
  Let $\vct{b}^\trans\vct{x}\le0$ be the triangular elimination
  of $\vct{a}^\trans\vct{x}\le0$.
  Let $f$ be the facet of $\CutC(G)$ supported by the inequality
  $\vct{a}^\trans\vct{x}\le0$,
  and $F$ be the face of $\CutC(G^+)$ supported
  by $\vct{b}^\trans\vct{x}\le0$.

  We prove the case of $a_{uu'}\le0$.
  The case of $a_{uu'}>0$ is proved by applying the case of $a_{uu'}\le0$
  to the switching of $f$ with respect to $\{u\}$.

  Let $\lvert E\rvert=d$ and $\lvert A\rvert=d'$.
  Because $f$ is a facet of $\CutC(G)$,
  there exist $d-1$ subsets $S_1,\dots,S_{d-1}\subseteq V\setminus\{u\}$
  such that $\delta_G(S_1),\dots,\delta_G(S_{d-1})$
  are linearly independent roots of $f$.

  Let $D=\{uu'\}\cup\{ij\mid i\in\{u,u'\},\;j\in A\}\subseteq E$,
  and let $\pi\colon\R^E\to\R^D$ be the orthogonal projection.
  Because there exists an edge $e\in E\setminus D$ such that $a_e\neq0$,
  the projected image $\pi(f)$ of $f$ to $\R^D$ is of full dimension
  by Lemma~\ref{lemm:projection}.
  This means that there exist $2d'+1$ subsets
  $T_1,\dots,T_{2d'+1}\subseteq V\setminus\{u\}$
  such that $\delta_G(T_1),\dots,\delta_G(T_{2d'+1})$ are roots of $f$
  and $\pi(\delta_G(T_1)),\dots,\pi(\delta_G(T_{2d'+1}))$
  are linearly independent.
  Note that the intersection of $\pi(\CutC(G))$
  and the hyperplane $x_{uu'}=0$ has a dimension $d'$,
  which means at most $d'$ out of $2d'+1$ subsets $T_1,\dots,T_{2d'+1}$
  satisfy $u'\notin T_i$.
  Therefore, at least $d'+1$ out of $2d'+1$ subsets $T_1,\dots,T_{2d'+1}$
  contain $u'$.
  Without loss of generality, we assume that
  $T_1,\dots,T_{d'+1}\ni u'$.
  For $1\le i\le d'+1$, let $T'_i=T_i\cup\{v\}$.

  Let
  \[
    C=\{\delta_{G^+}(S_1),\dots,\delta_{G^+}(S_{d-1}),
        \delta_{G^+}(T'_1),\dots,\delta_{G^+}(T'_{d'+1})\}.
  \]
  The $d+d'$ cut vectors in $C$ are roots of $F$.
  We prove that these $d+d'$ cut vectors are linearly independent.
  Let $M$ be a matrix of size $(d+d'+1)\times(d+d')$
  whose column vectors are these $d+d'$ cut vectors.
  We group the rows of $M$ into 4 groups $E_1$, $E_2$, $E_3$ and $E_4$, where
  \begin{align*}
    E_1&=E\setminus(E_2\cup E_3\cup E_4), \\
    E_2&=\{ui\mid i\in A\}, \\
    E_3&=\{vi\mid i\in A\}, \\
    E_4&=\{uv\}.
  \end{align*}
  Then $M$ is in the form
  \[
    M=\begin{matrix}
      \text{\footnotesize $E_1$} \\
      \text{\footnotesize $E_2$} \\
      \text{\footnotesize $E_3$} \\
      \text{\footnotesize $E_4$}
    \end{matrix}
    \begin{pmatrix}
      X & {*} \\
      Y & Z \\
      Y & \bm{1}-Z \\
      O & \bm{1}
    \end{pmatrix},
  \]
  where $\bm{1}$ represents a matrix whose elements are all $1$.
  The first $d-1$ columns of $M$ represent
  $\delta_{G^+}(S_1),\dots,\delta_{G^+}(S_{d-1})$
  and the last $d'+1$ columns represent
  $\delta_{G^+}(T'_1),\dots,\delta_{G^+}(T'_{d'+1})$.
  Because $\delta_{G^+}(S_1),\dots,\delta_{G^+}(S_{d-1})$
  are linearly independent,
  \[
    \rank\begin{pmatrix} X \\ Y \end{pmatrix}=d-1.
  \]
  Similarly, because
  $\pi(\delta_{G^+}(T'_1)),\dots,\pi(\delta_{G^+}(T'_{d'+1}))$
  are linearly independent,
  \[
    \rank\begin{pmatrix} Z \\ \bm{1}-Z \\ \bm{1} \end{pmatrix}=d'+1,
  \]
  which means
  \[
    \rank\begin{pmatrix} Z \\ \bm{1} \end{pmatrix}=d'+1.
  \]
  Therefore,
  \[
    \rank M=\rank\begin{pmatrix}
      X & {*} \\
      Y & Z \\
      O & -2Z \\
      O & \bm{1}
    \end{pmatrix}
    =\rank\begin{pmatrix} X \\ Y \end{pmatrix}
    +\rank\begin{pmatrix} -2Z \\ \bm{1} \end{pmatrix}
    =(d-1)+(d'+1)=d+d'.
  \]
  This means the $d+d'$ cut vectors in $C$
  are linearly independent roots of $F$,
  which means $F$ is a facet of $\CutC(G^+)$.
\end{proof}

Now we show that Theorems~\ref{thrm:valid} and \ref{thrm:facet-preserving} hold
also in the case of cut polytopes.
We will present two lemmas to establish the relation
between $\CutC(G)$ and $\CutP(G)$.
The first lemma contains well known facts (see, e.g.\ %
\cite[Section~26.3]{DezLau:cut97}).
We include the proof here for completeness.
Recall that the switching of the inequality $\vct{a}^\trans\vct{x}\le
a_0$ by the cut $\delta_G(S)$ is
\begin{equation}
  (\vct{a}^S)^\trans\vct{x}\le a_0-\vct{a}^\trans\delta_G(S).
  \label{eq:switching-0}
\end{equation}

\begin{lemma} \label{lemm:switch}
  Let $G=(V,E)$ be a graph,
  and $\vct{a}^\trans\vct{x}\le a_0$ be an inequality in $\R^E$.
  Let $S$ be a subset of $V$ such that the linear function
  $\vct{a}^\trans\vct{x}$ takes the maximum at $\delta_G(S)$ in $\CutP(G)$.
  Then the following conditions are equivalent.
  \begin{enumroman}
    \item
      The inequality $\vct{a}^\trans\vct{x}\le a_0$
      is valid for $\CutP(G)$.
    \item
      The switching of the inequality $\vct{a}^\trans\vct{x}\le a_0$
      by the cut $\delta_G(S)$ is valid for $\CutP(G)$.
    \item
      The switching of the inequality $\vct{a}^\trans\vct{x}\le a_0$
      by the cut $\delta_G(S)$ is valid for $\CutC(G)$.
  \end{enumroman}
  Similarly, the following conditions are equivalent.
  \begin{enumroman}
    \item
      The inequality $\vct{a}^\trans\vct{x}\le a_0$
      supports a facet of $\CutP(G)$.
    \item
      The switching of the inequality $\vct{a}^\trans\vct{x}\le a_0$
      by the cut $\delta_G(S)$ supports a facet of $\CutP(G)$.
    \item
      The switching of the inequality $\vct{a}^\trans\vct{x}\le a_0$
      by the cut $\delta_G(S)$ supports a facet of $\CutC(G)$.
  \end{enumroman}
\end{lemma}

\begin{proof}
  We first show the claim about the validity of the inequalities.
  The equivalence between (i) and (ii) is trivial
  because switching by any cut of $G$ maps $\CutP(G)$ onto itself.
  $\text{(iii)}\implies\text{(ii)}$ is also trivial because
  $\CutP(G)\subseteq\CutC(G)$.
  To show $\text{(ii)}\implies\text{(iii)}$,
  assume that \eqref{eq:switching-0} is valid for $\CutP(G)$.
  This means that for any $S'\subseteq V$,
  \begin{equation}
    (\vct{a}^S)^\trans\delta_G(S')\le a_0-\vct{a}^\trans\delta_G(S).
    \label{eq:switching-1}
  \end{equation}
  Letting $S'=\varnothing$ gives
  \begin{equation}
    a_0-\vct{a}^\trans\delta_G(S)\ge0.
    \label{eq:switching-2}
  \end{equation}
  By the definition of $S$ and the relation
  \[
    (\vct{a}^S)^\trans\delta_G(S')
    =\vct{a}^\trans\delta_G(S\triangle S')-\vct{a}^\trans\delta_G(S),
  \]
  the left hand side of \eqref{eq:switching-1} takes the maximum
  when $S\triangle S'=S$, or equivalently $S'=\varnothing$.
  This means
  \begin{equation}
    (\vct{a}^S)^\trans\delta_G(S')\le0
    \label{eq:switching-3}
  \end{equation}
  for any $S'\subseteq V$.
  Inequalities \eqref{eq:switching-2} and \eqref{eq:switching-3} give
  \[
    (\vct{a}^S)^\trans(\lambda\delta_G(S'))\le a_0-\vct{a}^\trans\delta_G(S)
  \]
  for any $S'\subseteq V$ and $\lambda\ge0$,
  which means \eqref{eq:switching-0} is valid for $\CutC(G)$.

  For the facet-supporting property, the argument is similar.
  $\text{(i)}\iff\text{(ii)}$ and $\text{(iii)}\implies\text{(ii)}$
  are trivial.
  To show $\text{(ii)}\implies\text{(iii)}$,
  assume that \eqref{eq:switching-0} supports a facet of $\CutP(G)$.
  From the argument above, especially \eqref{eq:switching-2} and
  \eqref{eq:switching-3}, it is necessary that
  $a_0-\vct{a}^\trans\delta_G(S)=0$
  for \eqref{eq:switching-0} to support a nonempty face of $\CutP(G)$.
  Because $\CutC(G)$ has every facet of $\CutP(G)$ that contains
  the coordinate origin, the inequality \eqref{eq:switching-0} supports
  a facet of $\CutC(G)$.
\end{proof}

\begin{lemma} \label{lemm:switch-tri-elim}
  Let $G=(V,E)$ be a graph,
  and let $G^+=(V^+,E^+)$ be the detour extension of $G$
  with removed edge $uu'\in E$, added vertex $v$ and adjacent vertex set $A$.
  Let $\vct{a}^\trans\vct{x}\le a_0$ be an inequality in $\R^E$,
  and $\vct{b}^\trans\vct{x}\le a_0$ be its triangular elimination.
  Then, there exists a subset $S$ of $V$
  such that the following conditions hold:
  \begin{enumroman}
    \item The switching of $\vct{b}^\trans\vct{x}\le a_0$
      by the cut $\delta_{G^+}(S)$ is the triangular elimination
      of the switching of $\vct{a}^\trans\vct{x}\le a_0$
      by the cut $\delta_G(S)$.
    \item The switching of $\vct{a}^\trans\vct{x}\le a_0$
      by the cut $\delta_G(S)$ is valid (resp.\ facet-supporting)
      for $\CutC(G)$
      if and only if $\vct{a}^\trans\vct{x}\le a_0$ is valid (resp.\ facet-supporting)
      for $\CutP(G)$.
    \item The switching of $\vct{b}^\trans\vct{x}\le a_0$
      by the cut $\delta_{G^+}(S)$ is valid (resp.\ facet-supporting)
      for $\CutC(G^+)$
      if and only if $\vct{b}^\trans\vct{x}\le a_0$ is valid (resp.\ facet-supporting)
      for $\CutP(G^+)$.
  \end{enumroman}
\end{lemma}

\begin{proof}
  First note that by definition,
  the inequality $\vct{b}^\trans\vct{x}\le a_0$ is written as
  \begin{equation}
    \vct{a}^\trans\vct{x}
      -a_{uu'}x_{uu'}+a_{uu'}x_{uv}-\lvert a_{uu'}\rvert x_{u'v}
      \le a_0.
    \label{eq:switch-tri-elim-2a}
  \end{equation}
  Let $S$ be a subset of $V\setminus\{u'\}$ such that
  the linear function $\vct{a}^\trans\vct{x}$ gives the maximum in $\CutP(G)$
  at the point $\delta(S)$.
  Then the switching of $\vct{a}^\trans\vct{x}\le a_0$
  by the cut $\delta_G(S)$ is
  \begin{equation}
    (\vct{a}^S)^\trans\vct{x}\le a_0-\vct{a}^S\delta_G(S),
    \label{eq:switch-tri-elim-3}
  \end{equation}
  and the switching of \eqref{eq:switch-tri-elim-2a}
  by the cut $\delta_{G^+}(S)$ is
  \begin{equation}
    (\vct{a}^S)^\trans\vct{x}
      -(-1)^{\chi_u(S)}a_{uu'}x_{uu'}
      +(-1)^{\chi_u(S)}a_{uu'}x_{uv}
      -\lvert a_{uu'}\rvert x_{u'v}
      \le a_0-\vct{a}^S\delta_G(S),
    \label{eq:switch-tri-elim-4}
  \end{equation}
  where $\chi_u(S)$ is $1$ if $u\in S$, or $0$ otherwise.
  We will check the conditions claimed in the lemma are satisfied.
  \begin{enumroman}
    \item By definition,
      the triangular elimination of \eqref{eq:switch-tri-elim-3} is
      the inequality \eqref{eq:switch-tri-elim-4}.
    \item This is proved by Lemma~\ref{lemm:switch}.
    \item Let $G'=(V^+,E^+\cup\{uu'\})$.
      Both $\vct{a}^\trans\vct{x}$ and
      $-a_{uu'}x_{uu'}+a_{uu'}x_{uv}-\lvert a_{uu'}\rvert x_{u'v}$
      are linear functions on $\R^{E^+\cup\{uu'\}}$.
      By definition of $S$, $\vct{a}^\trans\vct{x}$ takes the maximum
      at $\delta_{G'}(S)$ in $\CutP(G')$.
      In addition, the linear function
      $-a_{uu'}x_{uu'}+a_{uu'}x_{uv}-\lvert a_{uu'}\rvert x_{u'v}$
      takes the maximum value $0$ at $\delta_{G'}(S)$ in $\CutP(G')$.
      Therefore, the left hand side of \eqref{eq:switch-tri-elim-2a}
      takes the maximum at $\delta_{G'}(S)$ in $\CutP(G')$,
      which means it takes the maximum at $\delta_{G^+}(S)$ in $\CutP(G^+)$.
      Then the claim is proved by Lemma~\ref{lemm:switch}.  \qedhere
  \end{enumroman}
\end{proof}

\begin{theorem} \label{thrm:valid-polytope}
  Let $G=(V,E)$ be a graph,
  and let $G^+=(V^+,E^+)$ be the detour extension of $G$
  with removed edge $uu'\in E$, added vertex $v$ and adjacent vertex set $A$.
  Let $\vct{a}^\trans\vct{x}\le a_0$ be an inequality in $\R^E$,
  and $\vct{b}^\trans\vct{x}\le a_0$ be its triangular elimination.
  Then, the following two conditions are equivalent.
  \begin{enumroman}
    \item The inequality $\vct{a}^\trans\vct{x}\le a_0$ is valid
      for $\CutP(G)$.
    \item The inequality $\vct{b}^\trans\vct{x}\le a_0$ is valid
      for $\CutP(G^+)$.
  \end{enumroman}
\end{theorem}

\begin{proof}
  Let $S$ be the subset of $V$ stated in Lemma~\ref{lemm:switch-tri-elim}.
  Then condition (i) holds if and only if the switching of the inequality
  $\vct{a}^\trans\vct{x}\le a_0$ by the cut $\delta_G(S)$ is valid
  for $\CutC(G)$, and condition (ii) holds if and only if the switching of the inequality
  $\vct{b}^\trans\vct{x}\le a_0$ by the cut $\delta_{G^+}(S)$ is valid
  for $\CutC(G^+)$.
  By Theorem~\ref{thrm:valid}, the two conditions are equivalent.
\end{proof}

\begin{theorem} \label{thrm:facet-preserving-polytope}
  Let $G=(V,E)$ be a graph,
  and let $G^+=(V^+,E^+)$ be the detour extension of $G$
  with removed edge $uu'\in E$, added vertex $v$ and adjacent vertex set $A$.
  Let $\vct{a}^\trans\vct{x}\le a_0$ be an inequality
  supporting a facet of $\CutP(G)$.
  If there exists an edge $e\in E\setminus(\{uu'\}\cup\{uw,u'w\mid w\in A\})$
  such that $a_e\neq0$, then the triangular elimination of
  $\vct{a}^\trans\vct{x}\le a_0$ supports a facet of $\CutP(G^+)$.
\end{theorem}

\begin{proof}
  Let $\vct{b}^\trans\vct{x}\le a_0$ be the triangular elimination
  of $\vct{a}^\trans\vct{x}\le a_0$.
  Let $S$ be the subset of $V$ stated in Lemma~\ref{lemm:switch-tri-elim}.
  Then the switching of the inequality
  $\vct{a}^\trans\vct{x}\le a_0$ by the cut $\delta_G(S)$ supports a facet
  of $\CutC(G)$.
  By Theorem~\ref{thrm:facet-preserving},
  the switching of the inequality $\vct{b}^\trans\vct{x}\le a_0$
  by the cut $\delta_{G^+}(S)$
  supports a facet of $\CutC(G^+)$.
  This means the inequality $\vct{b}^\trans\vct{x}\le a_0$
  supports a facet of $\CutP(G^+)$.
\end{proof}

Note that in case of $a_{uu'}=-c<0$, we can consider, instead of
\eqref{eq:triangular-elimination}, the inequality
\begin{equation}
  \vct{a}^\trans\vct{x}
    +cx_{uu'}+cx_{uv}+cx_{u'v}\le a_0+2c,
  \label{eq:triangular-elimination-sw}
\end{equation}
which is obtained by adding a triangle inequality
$c(x_{uu'}+x_{uv}+x_{u'v})\le2c$
to the inequality $\vct{a}^\trans\vct{x}\le a_0$.
The inequality \eqref{eq:triangular-elimination-sw} is
the switching of the inequality \eqref{eq:triangular-elimination}
with respect to $\{v\}$.
This means two things:
\begin{itemize}
  \item The inequality \eqref{eq:triangular-elimination-sw} is valid
    for $\CutP(G^+)$ if and only if the inequality \eqref{eq:triangular-elimination}
    is valid for $\CutP(G^+)$.
  \item The inequality \eqref{eq:triangular-elimination-sw} supports a facet
    of $\CutP(G^+)$ if and only if the inequality \eqref{eq:triangular-elimination}
    supports a facet of $\CutP(G^+)$.
\end{itemize}

\subsection{Constructing facets of $\CutP(\K_{1,m,m})$
  by iterative triangular elimination}

Let $n>1$ be an integer and
$m = \frac{1}{2} \lfloor \frac{n-2}{2}\rfloor
                 \lfloor \frac{n-4}{2}\rfloor + n-2$.
The complete tripartite graph $\K_{1,m,m}$ is obtained
from the complete graph $\K_n$ by repeating the operation of
detour extension.
The next theorem follows from
Theorems~\ref{thrm:valid-polytope}
and \ref{thrm:facet-preserving-polytope}.

\begin{theorem}
  \label{thrm:kn-k1mm}
  Let $n>1$ be an integer and
  $m = \frac{1}{2} \lfloor \frac{n-2}{2}\rfloor
                   \lfloor \frac{n-4}{2}\rfloor + n-2$.
  \begin{enumroman}
    \item \label{enum:kn-k1mm-valid}
      Any valid inequality for $\CutP_n$
      can be converted to a valid inequality for $\CutP(\K_{1,m,m})$
      by repeating the operation of triangular elimination.
    \item \label{enum:kn-k1mm-facet}
      Any facet of $\CutP_n$ except for the triangle inequalities
      can be converted to a facet of $\CutP(\K_{1,m,m})$
      by repeating the operation of triangular elimination.
  \end{enumroman}
\end{theorem}

\begin{proof}
  \begin{enumroman}
    \item
      This is proved by applying Theorem~\ref{thrm:valid-polytope}
      repeatedly.
    \item
      For any facet of $\CutP_n$ except for the triangle inequalities,
      there exists at least four relevant vertices.
      Therefore, we can apply
      Theorem~\ref{thrm:facet-preserving-polytope} to every single
      operation of triangular elimination, and so we obtain the
      theorem.
  \end{enumroman}
\end{proof}

\begin{remark}
  As stated in Remark~\ref{remk:valid},
  any valid inequality for $\CutC(G)$ is converted to
  a valid inequality for $\CutC(G')$ by triangular elimination
  even if we do not put the condition $A\subseteq\nei_G(u)\cap\nei_G(u')$
  in the definition of detour extension.
  In this way, we can obtain a valid inequality for $\CutP(\K_{1,n-2,n-2})$
  from any valid inequality for $\CutP_n$.
  This ``compact'' construction allows much smaller $m$ for the same $n$ than
  Theorem~\ref{thrm:kn-k1mm}.
  In other words, this construction allows much greater $n$ for the same $m$,
  resulting in much more facets to apply this construction to.
  Without the condition on $A$,
  it can be proved that the dimension of the face obtained
  by triangular elimination is greater than or equal to the dimension
  of the original face.
  However, numerical tests shows that applying this conversion
  to most of the facets of $\CutP_n$ give faces
  of $\CutP(\K_{1,n-2,n-2})$ that are not facets.
  Because we are interested in computing facets of $\CutP(\K_{1,m,m})$,
  we stick to detour extension with the condition on $A$.
\end{remark}

\begin{example} \label{ex1}
  We show two facet-defining triangle inequalities for $\CutP_4$
  can be transformed into valid inequalities for $\CutP(\K_{1,2,2})$.
  First, consider the triangle inequality
  \begin{equation}
    x_{01} \le x_{02} + x_{12}. \label{eq1}
  \end{equation}
  The variable $x_{12}$ is not a valid variable for 
  $\CutP(\K_{1,2,2})$, so we eliminate it by adding \eqref{eq1} to
  the triangle inequality
  \[
    x_{12} \le x_{14} + x_{24}
  \]
  to obtain
  \begin{equation}
    x_{01} \le x_{02} + x_{14} + x_{24}. \label{eq3}
  \end{equation}
  By Theorem~\ref{thrm:kn-k1mm}~(\ref{enum:kn-k1mm-valid}),
  this is a valid inequality for $\CutP(\K_{1,2,2})$,
  but the face it supports is not a facet of $\CutP(\K_{1,2,2})$.
  The inequality \eqref{eq3} gives a valid inequality
  \[
    q_{(1,1),(2,2)}\le 2q_{(2,1),(2,1)}+q_{(2,1),(2,2)}
  \]
  for $\BellP(2,2,2)$.
  Next consider the triangle inequality
  \begin{equation}
    x_{12} \le x_{13} + x_{23}. \label{eq:ex1-5}
  \end{equation}
  We again eliminate $x_{12}$ adding
  \[
    x_{14} \le x_{12} + x_{24}
  \]
  to obtain
  \begin{equation}
    x_{14} \le x_{13} + x_{23} + x_{24}. \label{eq:example-tri-4}
  \end{equation}
  This time, \eqref{eq:example-tri-4} is not only valid
  for $\CutP(\K_{1,2,2})$,
  but it supports a facet of $\CutP(\K_{1,2,2})$.
  The inequality \eqref{eq3} gives a facet-defining inequality
\[
  q_{(2,2),(2,2)} \le q_{(1,2),(1,2)}+q_{(1,1),(2,2)} +q_{(2,2),(1,1)},
\]
  for $\BellP(2,2,2)$,
  which is known as the CHSH inequality~\cite{ClaHorShiHol-PRL69}.
\end{example}

\begin{example} \label{ex2}
  We show how a pentagon inequality, which defines a facet of $\CutP_5$
  can be transformed to a valid inequality for $\CutP(\K_{1,3,3})$.
  The pentagon inequality
  \[
    x_{01} + x_{12} + x_{02} + x_{45}
    \le x_{04} + x_{14} + x_{24}
     + x_{05} + x_{15} + x_{25}
  \]
  is transformed by adding to it the triangle inequalities
  \[
    x_{16} \le x_{12} + x_{26}
    \qquad\text{and}\qquad
    x_{34} \le x_{35} + x_{45}
  \]
  to give the inequality
  \begin{equation}
    x_{01} + x_{02} + x_{16} + x_{34}
    \le x_{04} + x_{14} + x_{24}
     + x_{05} + x_{15} + x_{25}
     +x_{26} + x_{35}.
    \label{eq6}
  \end{equation}
  This inequality gives the $\mathrm{I}_{3322}$ inequality~\cite{PitSvo-PRA01}
  for $\BellP(2,3,2)$.
\end{example}

\begin{example} \label{ex3}
  In Example~\ref{ex1},
  we considered two triangle inequalities.
  There is another case of triangle inequality
  which appears only in $\CutP_n$ for $n\ge6$.
  Let us consider the lifting of the triangle inequality
  \begin{equation}
    x_{13} \le x_{12} + x_{23}. \label{eq:ex3-1}
  \end{equation}
  for $\CutP_6$ to $\CutP(\K_{1,5,5})$.
  We have to eliminate all of the three variables in \eqref{eq:ex3-1}
  to obtain a valid inequality for $\CutP(\K_{1,5,5})$.
  To do this, we add the triangle inequalities
  \begin{align*}
    x_{17} &\le x_{13}+x_{37}, \\
    x_{12} &\le x_{18}+x_{28}, \\
    x_{23} &\le x_{29}+x_{39}
  \end{align*}
  to obtain a valid inequality
  \begin{equation}
    x_{17} \le x_{37} + x_{18} + x_{28}+x_{29}+x_{39} \label{eq:ex3-2}
  \end{equation}
  for $\CutP(\K_{1,5,5})$.
  The face it supports is not a facet of $\CutP(\K_{1,5,5})$.
\end{example}

\subsection{Equivalence of facets obtained by triangular elimination}
  \label{sbsect:equivalence}

Cut polytopes have many symmetries.
If we know one facet of $\CutP(\K_{1,m,m})$,
we can apply symmetric transformations to it
to obtain many different facets.
This leads to the question:
``How many different classes of facets of $\CutP(\K_{1,m,m})$ are obtained
by applying triangular elimination to facets of $\CutP_n$?''
In this subsection, we answer to this question
by establishing a relation between the equivalence of facets of $\CutP_n$
and the equivalence of their triangular eliminations.

\subsubsection{Definitions on symmetry of cut polytopes}

We need formal definitions to describe the symmetry of
$\CutP(\K_{1,m,m})$.

\paragraph*{Automorphism of graph}
An \emph{automorphism} of a graph $G=(V,E)$ is a permutation
$\sigma$ on $V$ such that
\[
  uv\in E \iff \sigma(u)\sigma(v)\in E.
\]
The set of all the automorphisms of $G$ is called the automorphism group
of $G$ and denoted by $\Aut(G)$.
For example, if $G$ is a complete graph $\K_n$,
then its automorphism group $\Aut(\K_n)$
is the symmetric group $\sym_n$ of degree $n$.

\medskip

In Section~\ref{sect:bell-cut} we introduced the switching operation,
which is one of the symmetries of the cut polytope.
Another is permutation.

\paragraph*{Permutation}
For a vector $\vct{x}\in\R^E$ and
an automorphism $\sigma\in\Aut(G)$ of $G$,
the \emph{permutation} of $\vct{x}$ by $\sigma$ is the vector
$\vct{x}'\in\R^E$ defined by
\[
  x'_{uv}=x_{\sigma(u)\sigma(v)},
\]
and denoted by $\sigma\cdot\vct{x}$.

For a vector $\vct{a}\in\R^E$, a scalar $a_0\in\R$
and an automorphism $\sigma\in\Aut(G)$,
the \emph{permutation} of the inequality $\vct{a}^\trans\vct{x}\le a_0$
by $\sigma$
is the inequality $(\sigma\cdot\vct{a})^\trans\vct{x}\le a_0$.
This permutation is valid for $\CutP(G)$
if and only if the original inequality $\vct{a}^\trans\vct{x}\le a_0$
is valid for $\CutP(G)$.
Similarly, the permutation supports a facet of $\CutP(G)$
if and only if the original inequality supports a facet of $\CutP(G)$.

For a facet $f$ of $\CutP(G)$ supported
by the inequality $\vct{a}^\trans\vct{x}\le a_0$,
the \emph{permutation} of $f$ by $\sigma$ is the facet of $\CutP(G)$
supported by the permutation of $\vct{a}^\trans\vct{x}\le a_0$ by $\sigma$,
and denoted by $\sigma\cdot f$.

\paragraph*{Facets of the same type}
Two facets $f$ and $f'$ of $\CutP(G)$ are \emph{switching equivalent},
denoted by $f\approx f'$,
if and only if there exists a set $S\subseteq V$ such that $\gamma(S)\cdot f=f'$.

Let $\calG$ be a subgroup of $\Aut(G)$.
Two facets $f$ and $f'$ of $\CutP(G)$ are \emph{$\calG$-permutation
equivalent}, denoted by $f\sim_\calG f'$,
if and only if there exists an automorphism $\sigma\in\calG$ of $G$
such that $\sigma\gamma(S)\cdot f\approx f'$.
In case of $\calG=\Aut(G)$, we say $f$ and $f'$ are \emph{of the same type}
instead of $\Aut(G)$-permutation equivalent, and denote this fact
by $f\sim f'$.

\paragraph*{Notation}
To keep the notations simple,
we focus on the cases where $n$ is odd
in most of the rest of this subsection.

Let $k$ be a natural number,
$n=2k+1$ and $m=k+\binom{k}{2}$.
Label the $n$ vertices of $\K_n=(V,E)$ by
$\X,\A_1,\dots,\A_k,\B_1,\dots,\B_k$,
and the $2m+1$ vertices of $\K_{1,m,m}$ by
\[
  \X;\A_1,\dots,\A_k,\A'_1,\dots,\A'_{\binom{k}{2}};
     \B_1,\dots,\B_k,\B'_1,\dots,\B'_{\binom{k}{2}}.
\]

Let $\langle{\cdot}\rangle\colon
     \binom{\{1,\dots,k\}}{2}\to\{1,\dots,\binom{k}{2}\}$ be a bijection.
Define the bijection
$\iota\colon\binom{\{\A_1,\dots,\A_k\}}{2}\cup\binom{\{\B_1,\dots,\B_k\}}{2}
 \to\{\A'_1,\dots,\A'_{\binom{k}{2}},\B'_1,\dots,\B'_{\binom{k}{2}}\}$ by
\begin{align*}
  \iota(\A_i\A_j)&=\B'_{\langle i,j\rangle}, \\
  \iota(\B_i\B_j)&=\A'_{\langle i,j\rangle}.
\end{align*}

\subsubsection{Switching equivalence}

\begin{theorem} \label{thrm:sw}
  Let $G=(V,E)$ be a graph,
  and let $G^+=(V^+,E^+)$ be the detour extension of $G$
  with removed edge $uu'\in E$, added vertex $v$ and adjacent vertex set $A$.
  Let $f$ and $f'$ be facets of $\CutP(G)$,
  and $F$ and $F'$ be the facets of $\CutP(G^+)$
  obtained as their triangular eliminations, respectively.
  Then,
  \[
    f\approx f' \iff F\approx F'.
  \]
\end{theorem}

\begin{proof}
  Let the inequality supporting $f$, $f'$, $F$ and $F'$ be
  \begin{align*}
    f &\colon\vct{a}^\trans\vct{x}\le a_0, \\
    f'&\colon\vct{a}^{\prime\trans}\vct{x}\le a'_0, \\
    F &\colon\vct{b}^\trans\vct{x}\le c_0, \\
    F'&\colon\vct{b}^{\prime\trans}\vct{x}\le c'_0,
  \end{align*}
  respectively.
  By the definition of triangular elimination,
  \begin{align*}
    \vct{b}^\trans\vct{x}&=\vct{a}^\trans\vct{x}
      -a_{uu'}x_{uu'}+a_{uu'}x_{uv}-\lvert a_{uu'}\rvert x_{u'v}, \\
    b_0&=a_0, \\
    \vct{b}^{\prime\trans}\vct{x}&=\vct{a}^{\prime\trans}\vct{x}
      -a'_{uu'}x_{uu'}+a'_{uu'}x_{uv}-\lvert a'_{uu'}\rvert x_{u'v}, \\
    b'_0&=a'_0.
  \end{align*}

  \noindent ($\Longrightarrow$) \quad
  Assume $f\approx f'$.
  Then there exists a set $S\subseteq V\setminus\{u'\}$ such that
  $f'=\gamma(S)\cdot f$.
  We will prove that $F'=\gamma(S)\cdot F$.

  Since $f'=\gamma(S)\cdot f$,
  we have $\vct{a}'=\vct{a}^S$ and $a'_0=a_0-\vct{a}^\trans\delta(S)$.
  Now it is sufficient if we prove $\vct{b}'=\vct{b}^S$.
  First, for $ww'\in E'$ such that $w,w'\neq v$, note that $ww'\in E$
  and
  \[
    b'_{ww'}=a'_{ww'}=a^S_{ww'}=b^S_{ww'}.
  \]
  Next,
  \begin{align*}
    b'_{uv}&=a'_{uu'}=a^S_{uu'}=\begin{cases}
      a_{uu'}=b_{uv}=b^S_{uv} & \text{if $u\notin S$,} \\
      -a_{uu'}=-b_{uv}=b^S_{uv} & \text{if $u\in S$,}
    \end{cases} \\
    b'_{u'v}&=-\lvert a'_{uu'}\rvert=-\lvert a^S_{uu'}\rvert
      =-\lvert a_{uu'}\rvert=b_{u'v}=b^S_{u'v}.
  \end{align*}
  Finally, for $w\in A$,
  we have $b_{vw}=b'_{vw}=0$ which means $b'_{vw}=b^S_{vw}$.
  Putting these equations together, we conclude that $\vct{b}'=\vct{b}^S$.

  \noindent ($\Longleftarrow$) \quad
  Assume $F\approx F'$.
  Then there exists a set $S'\subseteq V'\setminus\{u'\}$ such that
  $F'=\gamma(S')\cdot F$.
  Let $S=S'\setminus\{v\}$.
  Now we prove $f'=\gamma(S)\cdot f$.

  Since $F'=\gamma(S')\cdot F$,
  we have $\vct{b}'=\vct{b}^{S'}$ and
  $a'_0=a_0-\vct{b}^\trans\delta(S')=a_0-\vct{a}^\trans\delta(S)$.
  It is sufficient if we prove $\vct{a}'=\vct{a}^S$.

  For each $ww'\in E\setminus\{uu'\}$, we have $ww'\in E'$ and
  \[
    a'_{ww'}=b'_{ww'}=b^{S'}_{ww'}=a^S_{ww'}.
  \]
  In addition,
  \[
    a'_{uu'}=b'_{uv}=b^S_{uv}=\begin{cases}
      b_{uv}=a_{uu'}=a^S_{uu'} & \text{if $u\notin S'$,} \\
      -b_{uv}=-a_{uu'}=a^S_{uu'} & \text{if $u\in S'$,}
    \end{cases}
  \]
  and so $\vct{a}'=\vct{a}^S$.
\end{proof}

By applying Theorem~\ref{thrm:sw} repeatedly,
we obtain the following corollary.

\begin{corollary} \label{corl:sw}
  Let $f$ and $f'$ be facets of $\CutP_n$,
  and $F$ and $F'$ be the facets of $\CutP(\K_{1,m,m})$
  obtained by triangular elimination of $f$ and $f'$, respectively.
  Then,
  \[
    f\approx f' \iff F\approx F'.
  \]
\end{corollary}

\subsubsection{Switching permutation equivalence}

Here we consider the switching permutation equivalence of the facets
of $\CutP(\K_{1,m,m})$ obtained by triangular elimination of
facets of $\CutP_n$.

Note that in relation to Bell polytopes, the switching operation in
$\CutP(\K_{1,m,m})$ corresponds to the value exchange in
$\BellP(2,m,2)$, and the permutation operation in $\CutP(\K_{1,m,m})$
corresponds to the party and observable exchange in $\BellP(2,m,2)$.

Let $\calG_1$ be the subgroup of $\Aut(\K_n)$ generated by
\[
  \sym(\{\A_1,\dots,\A_k\})\cup\sym(\{\B_1,\dots,\B_k\}).
\]
Define $\sigma_0\in\Aut(\K_n)$ by
\[
  \sigma_0=\begin{pmatrix}
    \A_1 & \dotsm & \A_k & \B_1 & \dotsm & \B_k \\
    \B_1 & \dotsm & \B_k & \A_1 & \dotsm & \A_k
  \end{pmatrix},
\]
and let $\calG$ be the subgroup of $\Aut(\K_n)$
generated by $\calG_1\cup\{\sigma_0\}$.

\begin{theorem} \label{thrm:sw-perm}
  Let $n\ge5$ be an odd number.
  Let $f$ and $f'$ be non-triangle facets of $\CutP_n$,
  and $F$ and $F'$ be the facets of $\CutP(\K_{1,m,m})$
  obtained by triangular elimination of $f$ and $f'$, respectively.
  Then,
  \[
    f\sim_\calG f' \iff F\sim F'.
  \]
\end{theorem}

First, we consider the case of $k=1$.
In this case, $n=2k+1=3$ and $m=k+\binom{k}{2}=1$.
This means that $\CutP_n=\CutP_{1,m,m}$
and triangular elimination does nothing.
Theorem~\ref{thrm:sw-perm} is trivial in case of $k=1$.
In the rest of this section, we consider the case of $k>1$.

Let $\calH=\Aut(\K_{1,m,m})$,
and let $\calH_1$ be the subgroup of $\calH$ generated by
\[
  \sym(\{\A_1,\dots,\A_k,\A'_1,\dots,\A'_{\binom{k}{2}}\})\cup
  \sym(\{\B_1,\dots,\B_k,\B'_1,\dots,\B'_{\binom{k}{2}}\}).
\]
If we define $\tau_0\in\calH$ by
\[
  \tau_0=\left(\begin{array}{cccccccccccc}
    \A_1 & \dotsm & \A_k & \A'_1 & \dotsm & \A'_{\binom{k}{2}} &
    \B_1 & \dotsm & \B_k & \B'_1 & \dotsm & \B'_{\binom{k}{2}} \\
    \B_1 & \dotsm & \B_k & \B'_1 & \dotsm & \B'_{\binom{k}{2}} &
    \A_1 & \dotsm & \A_k & \A'_1 & \dotsm & \A'_{\binom{k}{2}}
  \end{array}\right),
\]
then $\calH$ is generated by $\calH_1\cup\{\tau_0\}$ since $k>1$.

For $\sigma\in\calG$, define $\tau\in\calH$ by
\[
  \tau(u)=\begin{cases}
    \sigma(u) &
      (\text{if $u\in\{\X,\A_1,\dots,\A_k,\B_1,\dots,\B_k\}$}), \\
    \iota(\sigma(v)\sigma(w)) &
      (\text{if $u=\iota(vw)$}).
  \end{cases}
\]
The mapping $\varphi\colon\calG\to\calH$
which maps each $\sigma\in\calG$ to $\tau\in\calH$ defined in this way
is a homomorphism between groups.
Let $\calH'=\im\varphi$ be the image of $\varphi$.

For now, we prove the following claim.

\begin{claim} \label{claim:sw-perm-1}
  $F\sim_\calH F'\implies F\sim_{\calH'}F'$.
\end{claim}

To prove this claim, we need some definitions to classify
the vertices of the graph $G$ according to their role
with respect to any given facet of $\CutP(G)$.

\begin{definition}[Irrelevant vertex of graph with respect to a facet]
  Let $G=(V,E)$ be a graph.
  A vertex $u\in V$ is \emph{irrelevant}
  with respect to a facet $f\colon\vct{a}^\trans\vct{x}\le a_0$ of $\CutP(G)$
  if and only if for any $v\in V$ such that $uv\in E$, we have $a_{uv}=0$.
\end{definition}

\begin{definition}[Triangular facet at vertex of graph]
  Let $G=(V,E)$ be a graph, and let $u\in V$.
  A facet $f\colon\vct{a}^\trans\vct{x}\le a_0$ is \emph{triangular} at $u$
  if and only if there exists two different vertices $v,v'\in V$ such that
  the following conditions are satisfied.
  \begin{enumroman}
    \item $uv,uv'\in E$, and $\lvert a_{uv}\rvert=\lvert a_{uv'}\rvert\neq0$.
    \item For any $w\in V\setminus\{v,v'\}$ such that $uw\in E$,
      we have $a_{uw}=0$.
  \end{enumroman}
  In such a case, we call the two vertices $v$ and $v'$
  the vertices \emph{adjacent} to $u$.
\end{definition}

\begin{lemma} \label{lemm:triangular-vertex}
  Let $f$ be a facet of $\CutP_n$.
  If $f$ is not a triangle inequality,
  then $f$ is not triangular at any vertex of $\K_n$.
\end{lemma}

\begin{proof}
  The proof is by contradiction.

  Let $\K_n=(V,E)$.
  Suppose a facet $f\colon\vct{a}^\trans\vct{x}\le a_0$ of $\CutP_n$
  is not a triangle inequality,
  and it is triangular at $u\in V$.
  Let $v,v'\in V$ be the vertices adjacent to $u$.
  Then $\lvert a_{uv}\rvert=\lvert a_{uv'}\rvert\neq0$.
  By switching $f$ by an appropriate subset of $\{u,v,v'\}$,
  we can assume $a_{uv}=a_{uv'}=-\lambda<0$ without loss of generality.

  A triangle inequality
  \begin{equation}
    \lambda(-x_{uv}-x_{uv'}+x_{vv'})\le0
    \label{eq:triangular-vertex-1}
  \end{equation}
  supports a facet of $\CutP_n$.
  Let $\vct{a}'\in\R^E$ be the vector which makes
  \[
    \vct{a}^{\prime\trans}\vct{x}
    =\vct{a}^\trans\vct{x}+\lambda(x_{uv}+x_{uv'}-x_{vv'})
  \]
  an identity.
  Since $f$ is not a triangle inequality, $\vct{a}'\neq0$.

  We prove the inequality $\vct{a}^{\prime\trans}\vct{x}\le a_0$ is
  valid for $\CutP_n$
  by showing that for any $S\subseteq V$,
  $\vct{a}^{\prime\trans}\delta(S)\le a_0$.
  Let $\vct{x}=\delta(S)$.
  Define $S'\subseteq V$ by
  \[
    S'=\begin{cases}
      S\cup\{u\}      & (\text{if $v\in S$,}) \\
      S\setminus\{u\} & (\text{if $v\notin S$,})
    \end{cases}
  \]
  and let $\vct{x}'=\delta(S')$.
  Since $f$ is triangular at $u$ adjacent to $v$ and $v'$,
  the inequality $\vct{a}^{\prime\trans}\vct{x}\le a_0$ does not have
  any terms that correspond to edges incident to $u$,
  which means $\vct{a}^{\prime\trans}\vct{x}=\vct{a}^{\prime\trans}\vct{x}'$.
  Because $x'_{uv}=x'_{uv'}-x'_{vv'}=0$,
  \[
    \vct{a}^{\prime\trans}\vct{x}
    =\vct{a}^{\prime\trans}\vct{x}'
    =\vct{a}^\trans\vct{x}'+\lambda(x'_{uv}+x'_{uv'}-x'_{vv'})
    =\vct{a}^\trans\vct{x}'\le a_0,
  \]
  which means the inequality $\vct{a}^{\prime\trans}\vct{x}\le a_0$
  is valid for $\CutP_n$.

  The inequality $\vct{a}^\trans\vct{x}\le a_0$
  is the sum of the inequality~\eqref{eq:triangular-vertex-1}
  and the inequality $\vct{a}^{\prime\trans}\vct{x}\le a_0$.
  This means that $f$ is not a facet,
  hence a contradiction.
  Therefore, $f$ is not triangular at any vertex $u$.
\end{proof}

Claim~\ref{claim:sw-perm-1} is proved
from Lemma~\ref{lemm:triangular-vertex} as follows.

\begin{proof}[Proof of Claim~\ref{claim:sw-perm-1}]
  Before the main part of the proof,
  consider the case when $F$ is a triangle inequality.
  By $F\sim_{\calH}F'$, $F'$ is also a triangle inequality.
  Then $F\approx F'$, which trivially implies $F\sim_{\calH'}F'$.

  Now assume that $F$ is not a triangle inequality.
  By $F\sim_{\calH}F'$, $F'$ is not a triangle inequality.
  First, we will prove the case of $\tau\in\calH_1$.

  Let $\K_n=(V,E)$.
  We classify the elements of $V$ by their roles in the facet $F$
  into three groups:
  \begin{align*}
    V_1&=\{u\in V\mid\text{$u$ is irrelevant with respect to $F$}\}, \\
    V_2&=\{u\in V\mid\text{$F$ is triangular at $u$}\}, \\
    V_3&=V\setminus(V_1\cup V_2).
  \end{align*}
  Note that $V=V_1\cup V_2\cup V_3$ is a partition of $V$
  into disjoint union.
  Similarly, let
  \begin{align*}
    V'_1&=\{u\in V\mid\text{$u$ is irrelevant with respect to $F'$}\}, \\
    V'_2&=\{u\in V\mid\text{$F'$ is triangular at $u$}\}, \\
    V'_3&=V\setminus(V'_1\cup V'_2).
  \end{align*}
  Since $F'\approx\tau\cdot F$,
  \[
    \tau(V_1)=V'_1,\quad\tau(V_2)=V'_2,\quad\tau(V_3)=V'_3.
  \]
  By Lemma~\ref{lemm:triangular-vertex},
  \begin{equation}
    V_2,V'_2\subseteq\{\A'_1,\dots,\A'_{\binom{k}{2}},
                       \B'_1,\dots,\B'_{\binom{k}{2}}\}.
    \label{eq:sw-perm-claim-1-1}
  \end{equation}
  By the definition of triangular elimination,
  \begin{align*}
    \{\A'_1,\dots,\A'_{\binom{k}{2}},
      \B'_1,\dots,\B'_{\binom{k}{2}}\}&\subseteq V_1\cup V_2, \\
    \{\A'_1,\dots,\A'_{\binom{k}{2}},
      \B'_1,\dots,\B'_{\binom{k}{2}}\}&\subseteq V'_1\cup V'_2,
  \end{align*}
  which means
  \begin{equation}
    V_3,V'_3\subseteq\{\A_1,\dots,\A_k,\B_1,\dots,\B_k\}.
    \label{eq:sw-perm-claim-1-2}
  \end{equation}
  From the relations~\eqref{eq:sw-perm-claim-1-1} and
  \eqref{eq:sw-perm-claim-1-2},
  there exist two permutations $\sigma\in\calG_1$ and
  $\bar{\sigma}\in\sym(\{\A'_1,\dots,\A'_{\binom{k}{2}}\})\times
                  \sym(\{\B'_1,\dots,\B'_{\binom{k}{2}}\})$ such that
  \begin{alignat*}{2}
    \sigma(u)&=\tau(u) &\quad &(\forall u\in V_3), \\
    \bar{\sigma}(u)&=\tau(u) &\quad &(\forall u\in V_2).
  \end{alignat*}
  Since $(\sigma\bar{\sigma})(u)=\tau(u)$ for any $u\in V\setminus V_1$,
  $F'\approx\sigma\bar{\sigma}\cdot F$.

  By comparing the coefficients of $F'$ and $\varphi(\sigma)\cdot F$
  in a similar way to the proof of Lemma~\ref{lemm:triangular-vertex},
  we have $F'\approx\varphi(\sigma)\cdot F$.
  This completes the proof in case of $\tau\in\calH_1$.

  In case of $\tau\notin\calH_1$,
  $\tau$ can be written as $\tau=\tau'\tau_0$ by using some $\tau'\in\calH_1$.
  Note that
  \[
    F'\approx\tau\cdot F=\tau'\tau_0\cdot F.
  \]
  From what we already proved, there exists $\sigma'\in\calG$
  such that $F'\approx\varphi(\sigma')\tau_0\cdot F$.
  Since $\tau_0=\varphi(\sigma_0)$,
  we have $F'\approx\varphi(\sigma'\sigma_0)\cdot F$.
  This means $F\sim_{\calH'}F'$ in case of $\tau\notin\calH_1$.
\end{proof}

The following claim is straightforward from the definition of
triangular elimination.

\begin{claim} \label{claim:sw-perm-2}
  For $\sigma\in\calG$, \quad
  $f'\approx\sigma\cdot f\iff F'\approx\varphi(\sigma)\cdot F$.
\end{claim}

Theorem~\ref{thrm:sw-perm} immediately follows
Claims~\ref{claim:sw-perm-1} and \ref{claim:sw-perm-2}.
It answers the question we posed at the beginning
of Section~\ref{sbsect:equivalence} in cases where $n$ is odd.
For example, $\CutP_7$ has 67 different classes
of $\calG$-permutation equivalent facets,
where 4 out of them are triangle inequalities:
\begin{itemize}
  \item
    $x_{\X\A_1}-x_{\X\B_1}-x_{\A_1\B_1}\le0$,
    which is itself a facet-supporting inequality
    of $\CutP(\K_{1,m,m})$ and corresponds to the trivial inequality of
    $\BellP(2,m,2)$.
  \item
    $x_{\A_1\A_2}-x_{\A_1\B_1}-x_{\A_2\B_1}\le0$,
    like \eqref{eq:ex1-5} in Example~\ref{ex1},
    whose triangular elimination gives a facet-supporting inequality
    of $\CutP(\K_{1,m,m})$ and corresponds to the CHSH inequality of
    $\BellP(2,m,2)$.
  \item
    $x_{\X\A_1}-x_{\X\A_2}-x_{\A_1\A_2}\le0$,
    like \eqref{eq1} in Example~\ref{ex1},
    whose triangular elimination does not support a facet of
    $\CutP(\K_{1,m,m})$.
  \item
    $x_{\A_1\A_3}-x_{\A_1\A_2}-x_{\A_2\A_3}\le0$,
    like \eqref{eq:ex3-1} in Example~\ref{ex3},
    whose triangular elimination does not support a facet of
    $\CutP(\K_{1,m,m})$.
\end{itemize}
This means $\CutP(\K_{1,6,6})$ has $63$ different classes
of facets of the same type which can be obtained
by applying triangular elimination to non-triangular facets of $\CutP_7$.

In cases where $n$ is even, we need special care to define what corresponds
to the subgroup $\calG$ of $\Aut(\K_n)$.
Let $n=2k$ and label the $n$ vertices of $\K_n$
by $\X,\A_1,\dots,\A_k,\B_1,\dots,\B_{k-1}$.
We can define $\calH_1$ and $\tau_0$ in the same way as the cases
where $n$ is odd, and $\calH_1\cup\{\tau_0\}$ generates the group
$\Aut(\K_{1,m,m})$.
The problem is that when $n$ is even,
there does not exist $\sigma_0\in\Aut(\K_n)$
such that $\varphi(\sigma_0)=\tau_0$.
Therefore, we take a different approach.
We regard $\K_n$ as a subgraph of $\K_{n+1}$
which has an extra vertex $\B_k$.
For any facet $f$ of $\CutP_n$, its 0-lifting $\bar{f}$
is a facet of $\CutP_{n+1}$ by the 0-lifting theorem~\cite{Des-ORL90}.
We say two facets $f$ and $f'$ of $\CutP_n$ are \emph{equivalent}
if and only if their 0-lifting $\bar{f}$ and $\bar{f}'$ satisfy
$\bar{f}\sim_\calG\bar{f}'$, where $\calG$
is the subgroup of $\Aut(\K_{n+1})$ defined above.
Let $F$ and $F'$ be the facet of $\CutP(\K_{1,m,m})$ obtained
by applying triangular elimination to $f$ and $f'$, respectively.
Similarly, let $\bar{F}$ and $\bar{F}'$
be the facet of $\CutP(\K_{1,m+1,m+1})$ obtained by applying
triangular elimination to $\bar{f}$ and $\bar{f}'$, respectively.
Then $\bar{F}$ and $\bar{F}'$ is the 0-lifting of $F$ and $F'$.
This means that $F$ and $F'$ are of the same type
if and only if $\bar{F}$ and $\bar{F}'$ are of the same type.
Therefore, the following fact holds.

\begin{corollary} \label{corl:sw-perm-n-even}
  Let $n\ge4$ be an even number.
  Let $f$ and $f'$ be non-triangle facets of $\CutP_n$,
  and $F$ and $F'$ be the facets of $\CutP(\K_{1,m,m})$
  obtained by triangular elimination of $f$ and $f'$, respectively.
  Then,
  \[
    \text{$f$ and $f'$ are equivalent} \iff F\sim F'.
  \]
\end{corollary}

\begin{table}
  \centering
  \caption{The number $C_n$ of the classes of facets of $\CutP(\K_{1,m,m})$
    of the same type obtained by applying triangular elimination
    to non-triangular facets of $\CutP_n$.
    The values of $C_8$ and $C_9$ depend on the conjecture
    that the lists of facets of $\CutC_8$ and $\CutC_9$
    on the Web site~\cite{Smapo} are complete.}
  \label{table:sw-perm}
  \begin{tabular}{|c|c|c|c|c|c|} \hline
     $n$  & 5 & 6 &  7 &      8 &          9 \\ \hline
     $m$  & 3 & 5 &  6 &      9 &         10 \\ \hline
    $C_n$ & 1 & 6 & 63 & 16,234 & 44,368,793 \\ \hline
  \end{tabular}
\end{table}

By Theorem~\ref{thrm:sw-perm} and Corollary~\ref{corl:sw-perm-n-even},
we can compute the number of the classes of facets of $\CutP(\K_{1,m,m})$
of the same type obtained by applying triangular elimination
to non-triangular facets of $\CutP_n$.
We consulted De~Simone, Deza and Laurent~\cite{DesDezLau-DM94}
for the H-representation of $\CutC_7$,
and the ``conjectured complete description'' of $\CutC_8$
and the ``description possibly complete'' of $\CutC_9$ in SMAPO~\cite{Smapo}.
The result is summarized in Table~\ref{table:sw-perm}.

\section{Tight Bell inequalities from triangular elimination}
As stated in previous section, triangular elimination preserves facet
supporting property
(Theorem~\ref{thrm:kn-k1mm}~(\ref{enum:kn-k1mm-facet})) and
inequivalence property under known isomorphisms
(Corollary~\ref{corl:sw-perm-n-even}), which correspond to party,
observable and value exchanges.
As a consequence, we can obtain a large number of tight Bell inequalities.

In this section, we compile the results of triangular elimination in
the form of Bell inequalities. Throughout the rest of this section,
we use the term ``family'' as set of Bell inequalities, on the other
hand, the term ``class'' as set of facets of cut polytope.
In addition, we denote ${q}_{(j,2),(j^{\prime},2)}$ and
${q}_{(j,2),(j^{\prime},2)} + {q}_{(j,2),(j^{\prime},1)}$ as
$q_{A_{j}B_{j^{\prime}}}$ and ${q}_{A_{j}}$ respectively, and define
$q_{B_{j^{\prime}}}$ similarly. Then, terms of the left
hand side of inequality are arrayed in the format introduced by
Collins and Gisin~\cite{ColGis-JPA04}; each row corresponds to
coefficients of each observable of party $A$ and each column
corresponds to that of party $B$.  Because of switching equivalence,
we can assume that the right hand side of inequality are always zero.
The example of the CHSH $q_{(1,2),(1,2)} - q_{(1,2),(2,1)} -
q_{(2,1),(1,2)} - q_{(2,2),(2,2)} = -q_{A_{1}} -q_{B_{1}} +
q_{A_{1}B_{1}} + q_{A_{1}B_{2}} + q_{A_{2}B_{1}} - q_{A_{2}B_{2}} \leq 0$
is arrayed in the form as follows:
\begin{align}
	\left( \begin{array}{c||cc}
	  &-1 & \\ \hline \hline
	-1& 1 & 1\\
	  & 1 &-1
	\end{array} \right).
\end{align}

Note that the complete graph $K_{2k+1}$ for some $k$ has symmetric group
$\mathcal{S}_{2k+1}$ as its automorphism group, on the other hand, the
complete tripartite graph $K_{1,k,k}$ has only the subgroup of
$\mathcal{S}_{2k+1}$. Therefore, some classes of facets which are
originally equivalent as a facet class of $\CutP_{2k+1}$ under
permutation and switching can define inequivalent families of Bell
inequalities.

\subsection{Family of tight Bell inequalities obtained from triangular elimination of  hypermetric facet}

In the case of cut polytope of complete graph, some explicit classes
of valid inequalities are known, for example, the hypermetric,
clique-web and gap inequalities~\cite[Part V]{DezLau:cut97}.  For
these classes of inequalities, some sufficient conditions to be facet
supporting are also known. Therefore through triangular elimination,
we can obtain families of tight Bell inequalities from them.
 
 First, we give a new family of tight Bell
 inequalities found by applying triangular elimination to
 the hypermetric inequality class~\cite[Chapter~28]{DezLau:cut97}.
 A special case of this
 family, namely the triangular eliminated pure hypermetric inequality,
 contains four previously known Bell inequalities: the trivial
 inequalities like $q_{A_{1}} \leq 1$, the well
known CHSH inequality found by Clauser, Horne, Shimony and
Holt~\cite{ClaHorShiHol-PRL69}, the inequality named $I_{3322}$ by
Collins and Gisin~\cite{ColGis-JPA04},
originally found by Pitowsky
and Svozil~\cite{PitSvo-PRA01}, and the $I_{3422}^2$ inequality by
Collins and Gisin~\cite{ColGis-JPA04}.

Let $n = 2k+1$ for some $k$, $\vct{a} \in \Z^{\{A_{1} ,\ldots, A_{k}\}}, \vct{b} \in
\Z^{\{B_{1} ,\ldots, B_{k}\} }$ be integer weight vectors
for each observable and $c \in
\Z$ satisfying $ c + \sum_{j=1}^{k} a_{A_{j}} + \sum_{j^{\prime}=1}^{k} b_{B_{j}} = 1$.
Because of equivalence under observable exchange,
we can assume that without loss of generality, the elements of
$\vct{a},\vct{b}$ are sorted in some manner. Similarly, exchange of
$\vct{a}$ and $\vct{b}$ does not yields new family.

Then the following inequality is always a valid Bell inequality:
\vspace{-\baselineskip}\vspace{4pt}
\begin{multline}
   \sum_{j = 1}^{k}(1 - a_{A_{j}} - 2 \sum_{j^{\prime}=1}^{j-1} a_{A_{j^{\prime}}} ) a_{A_{j}}q_{A_{j}}
 + \sum_{j = 1}^{k}(1 - b_{B_{j}} - 2 \sum_{j^{\prime}=1}^{j-1} b_{B_{j^{\prime}}} ) b_{B_{j}}q_{B_{j}}\\
   - 2 \sum_{1 \leq j,j^{\prime} \leq k} a_{A_{j}} b_{B_{j^{\prime}}} q_{A_{j}B_{j^{\prime}}}\\
   - 2 \sum_{uu^{\prime} \in \binom{k}{2}} a_{A_{u}}a_{A_{u^{\prime}}} (q_{A_{u}B_{v_{uu^{\prime}}}} - q_{A_{u^{\prime}}B_{v_{uu^{\prime}}}})
   - 2 \sum_{uu^{\prime} \in \binom{k}{2}} b_{B_{u}}b_{B_{u^{\prime}}} (q_{A_{v_{uu^{\prime}}}B_{u}} - q_{A_{v_{uu^{\prime}}}B_{u^{\prime}}}) \leq 0,
  \label{eq:tri-elim-hypermetric}
\end{multline}
where $v_{uu^{\prime}}$ is the added vertex in each step of detour extension corresponding to $uu^{\prime}$.

If $\vct{a}$, $\vct{b}$ and $c$ satisfy one of the following
conditions in addition to the condition
$c+\sum_{j=1}^k a_{A_j}+\sum_{j'=1}^k b_{B_{j'}}=1$ stated above, then
the corresponding hypermetric inequality is
facet-supporting~\cite[Corollary~28.2.5]{DezLau:cut97}, which gives
that the inequality~\eqref{eq:tri-elim-hypermetric} is a tight Bell
inequality.
\begin{enumerate}
  \item \label{enum:pure-hypermetric}
    Each of $a_{A_j}$, $b_{B_{j'}}$ and $c$ is $1$, $-1$ or $0$.
    (In this case, the corresponding hypermetric inequality is called
     a pure hypermetric inequality, or a pure $l$-gonal inequality if
     there are $l$ nonzero elements in $a_{A_j}$, $b_{B_{j'}}$ and
     $c$.)
  \item
    In $a_{A_j}$, $b_{B_{j'}}$ and $c$, there are at least 3 and at
    most $n-3$ positive elements and there are no elements less than
    $-1$.
\end{enumerate}

Let us look at the case~\ref{enum:pure-hypermetric} more closely.
If we let $k=1$ and $(\vct{a}^\trans,\vct{b}^\trans,c)=((1),(-1),1)$,
we obtain the trivial inequality.
If we let $k=2$ and
$(\vct{a}^\trans,\vct{b}^\trans,c)=((1,1),(-1,-1),1)$, we obtain the
$I_{3322}$ inequality (affinely isomorphic to inequality~\eqref{eq6}
in Example~\ref{ex2}).
If we let $k=2$ and
$(\vct{a}^\trans,\vct{b}^\trans,c)=((1,1),(0,-1),-1)$, we obtain the
CHSH inequality (affinely isomorphic to
inequality~\eqref{eq:example-tri-4} in Example~\ref{ex1}).

If we let $k=3$ and
$(\vct{a}^\trans,\vct{b}^\trans,c)=((-1,1,1),(-1,0,0),1)$, we obtain
the $I^{2}_{3422}$ inequality found by Collins and
Gisin~\cite{ColGis-JPA04}, a tight Bell inequality for an asymmetric
2-party setting in which one party has 3 2-valued observables and
the other party has 4 2-valued observables (the right hand side is
explicitly given here because it is not zero):
\begin{equation}
I^{2}_{3422} =
	\left( \begin{array}{c||ccc}
  &  & 1&-1\\ \hline \hline
-1&-1& 1& 1\\
  &  &-1& 1\\
-1& 1&  & 1\\
 1&-1&-1& 
	\end{array}\right)\le1
\label{eq:I3422-2}
\end{equation}

By assigning $\vct{a}$, $\vct{b}$ and $c$ satisfying the condition
mentioned above, infinitely many tight Bell inequalities are obtained.
For example, by letting $k=3$ and
$(\vct{a}^\trans,\vct{b}^\trans,c)=((1,1,1),(-1,-1,-1),1)$,
$\BellP(2,6,2)$ has the following facet:
\begin{align}
	\left( \begin{array}{c||ccc:ccc}
   &-1&-2&-3&  &  & \\ \hline \hline
   & 1& 1& 1&-1&-1& \\
 -1& 1& 1& 1& 1&  &-1\\
 -2& 1& 1& 1&  & 1& 1\\ \hdashline
   &-1& 1&  &  &  & \\
   &-1&  & 1&  &  & \\
   &  &-1&-1&  &  & \\
	\end{array} \right).
\end{align}

\subsection{Other families of tight Bell inequality}
There are more general classes of facets in cut polytope of the complete
graph. Of these classes, the clique-web inequalities contains
hypermetric inequalities as a special case. There are also known
sufficient conditions for clique-web inequalities be facet
supporting.

For example, the \emph{pure} clique-web inequality is facet supporting~\cite[Section~29.4]{DezLau:cut97}.
Using triangular elimination, for $m \geq 7$ we can also obtain families of tight Bell inequalities like
\begin{align}
	\left( \begin{array}{c||ccccccc}
  &  &  &  &-1&-2&-2&-2\\ \hline \hline
-3& 1& 1& 1& 1& 1& 1& 1\\ \hdashline
  &-1&  &  & 1&  &  & \\
  &-1&  &  &  & 1&  & \\
  &  &-1&  &  & 1&  & \\
  &  &-1&  &  &  & 1& \\
  &  &  &-1&  &  & 1& \\
  &  &  &-1&  &  &  & 1\\
  &  &  &  &-1&  &  & 1\\
	\end{array}\right).
\end{align}
Note that because of the complexity of the structure, there are large
number of triangular eliminated clique-web facets which are equivalent
in the $\CutP_{n}$ but not equivalent as families of tight Bell
inequalities. These separated families are induced by the original
classes and the embedding into the $\CutP(K_{1,m,m})$.

\subsection{Relationship between $I_{mm22}$ and triangular eliminated Bell inequality}
Collins and Gisin~\cite{ColGis-JPA04} proposed a family of tight Bell
inequalities obtained by the extension of CHSH and $I_{3322}$ as
$I_{mm22}$ family, and conjectured that $I_{mm22}$ is always facet
supporting (they also confirmed that for $m \leq 7$, $I_{mm22}$ is
actually facet supporting by computation). Therefore, whether their
$I_{mm22}$ can be obtained by triangular elimination of some facet
class of $\CutP_{n}$ is an interesting question.

The $I_{mm22}$ family has the structure as follows:

\begin{align}
	\left( \begin{array}{c||ccccc}
      &-1    &       &  &  &\\ \hline \hline
-(m-1)& 1    &\cdots & 1& 1& 1\\
-(m-2)&      &\cdots & 1& 1&-1\\
-(m-3)& 1    &\cdots & 1&-1&  \\
\vdots&\vdots&\cdots &\revddots&  &\\
    0 & 1    &-1&                 &                 &
	\end{array} \right).
\end{align}

From its structure, it is straightforward that if $I_{mm22}$ can be obtained by
triangular elimination of some facet class of $\CutP_{n}$, then only
$A_{m}$ and $B_{m}$ are detour vertices,
since other vertices have degree more than $2$. However, there is no
known corresponding facet class of cut polytope in this form. 
For specific values, by computation, we found that corresponding facets for
$m=2,3,4$ are the triangle, pentagonal and Grishukhin~\cite{Gri-EJC90}
inequality $\sum_{1 \leq i < j \leq 4}x_{ij} +x_{56} +x_{57} -x_{67} -x_{16} -x_{36}
-x_{27} -x_{47} - 2 \sum_{1 \leq i \leq 4}x_{i5} \leq 0$, respectively.

\subsection{Facet of $\BellP(2,m,2)$ other than the triangular elimination of $\CutP_{n}$}
Since we have obtained a large number of tight Bell inequalities by
triangular elimination of $\CutP_{n}$, the next question is whether they
are complete i.e., whether all families and its equivalents form
whole set of facets of $\BellP(2,m,2)$. 

For $m=3$ case, this is affirmative.  Both \'{S}liwa~\cite{Sli-PLA03}
and Collins and Gisin~\cite{ColGis-JPA04} showed that there are only
three kinds of inequivalent facets: the trivial, CHSH and $I_{3322}$,
corresponding to the triangle facet, the triangular elimination of the
triangle facet and the triangular elimination of the pentagonal facet
of $\CutP_{n}$, respectively.

On the other hand, for $m \geq 4$, the answer is negative because there is
facet, found by facet enumeration of $\BellP(2,4,2)$ by
lrs~\cite{Avi:lrs}, such as
\begin{align}
	\left( \begin{array}{c||cccc}
  &  &-1& 1&-1\\ \hline \hline
  & 2& 1& 1& \\
 1&  & 1&-1&-1\\
 1& 1&  &-1& 1\\
 1&-1& 1&  & 1
	\end{array}\right).
\end{align}
The counterpart of this inequality in cut polytope is neither a facet of
$\CutP_{n}$ nor the triangular elimination of any facet of
$\CutP_{n}$ because it has no vertex with degree $2$.

Actually, in the asymmetric setting with 2 parties having 3 and 4
2-valued observables, Collins and Gisin enumerated all of the tight
Bell inequalities
and classified them into 6 families of equivalent inequalities~\cite{ColGis-JPA04}.
While the trivial, CHSH, $I_{3322}$ and $I_{3422}^2$ inequalities are
either facets of $\CutP_n$ or their triangular eliminations, the other
two are not:
\begin{align}
I^{1}_{3422} =
	\left( \begin{array}{c||ccc}
  & 1& 1&-2\\ \hline \hline
 1&-1&-1& 1\\
  &-1& 1& 1\\
  & 1&-1& 1\\
 1&-1&-1&-1
	\end{array}\right)\le2, \quad
I^{3}_{3422} =
	\left( \begin{array}{c||ccc}
  & 1&  &-1\\ \hline \hline
  &-2& 1& 1\\
  &  &-1& 1\\
-1& 1& 1& 1\\
 2&-1&-1&-1 
	\end{array}\right)\le2.
\label{eq:I3422-13}
\end{align}
We also consider the inequalities~\eqref{eq:I3422-2} and
\eqref{eq:I3422-13} later in Section~\ref{sec:asymm}.

\section{$\NP$-completeness of membership testing}
  \label{sect:intractable}

In this section, we consider the computational complexity of the problem
to determine whether a given correlation table $\vct{q}\in\R^{(mv)^n}$
is induced by some classical $(n,m,v)$-system or not.
This problem is rephrased in polytopal terminology as follows.

\begin{quote}
  \begin{problem}
    \item Membership test for Bell polytope
    \item Instance: A positive integer $m$ and a vector
          $\vct{q}\in\Q^{(2m)^2}$.
    \item Question: Is $\vct{q}\in\BellP(2,m,2)$?
  \end{problem}
\end{quote}

\begin{theorem} \label{thrm:bell-hard}
  The membership test for the Bell polytope is
  $\NP$-complete in the sense of polynomial-time Turing reducibility%
  \footnote{
    Pitowsky~\cite{Pit-MP91} shows that the membership test
    for correlation polytope of a given graph $G$
    is $\NP$-complete in the sense of polynomial-time Karp reducibility
    by reduction from one-in-three SAT.
  }.
\end{theorem}

The membership test for Bell polytope is in $\NP$
by Carath\'{e}odory's theorem.
The proof of $\NP$-hardness can be sketched as follows.
First, we prove the affine isomorphism between Bell polytopes $\BellP(2,m,2)$
and cut polytopes of tri-partite graphs $\K_{1,m,m}$.
Next, we prove the weighted maximum cut problem on $\K_{1,m,m}$ is
$\NP$-complete.
Finally, we prove the weighted maximum cut problem on $\K_{1,m,m}$
is polynomial-time Turing reducible to the membership test of
cut polytopes of $\K_{1,m,m}$
in a similar way to the proof of the $\NP$-hardness of the membership test
of cut polytopes of complete graphs.

\subsection{$\NP$-completeness of weighted maximum cut on $\K_{1,m,m}$}

\begin{quote}
  \begin{problem}
    \item Weighted maximum cut
    \item Instance: A graph $G=(V,E)$,
          an integer vector $\vct{w}\in\Z^E$,
          and an integer $k$.
    \item Question: Is there a subset $C\subseteq V$
          that satisfies the condition
          \[
            \sum_{\substack{u\in C \\ v\in V\setminus C}}w_{uv}\ge k?
          \]
  \end{problem}
\end{quote}

Weighted maximum cut is $\NP$-complete.
See Garey and Johnson~\cite{GarJoh:computers79} for a proof.

Let $G=(V,E)$ be a graph and $H=(V',E')$ be a minor of $G$.
Weighted maximum cut on $H$ is easier that on $G$
because we can reduce the former to the latter by assigning
0 to removed edges and $-M$ to contracted edges,
where $M$ is any integer greater than $\sum_{e\in E'}t_e$.

Any graph with $m$ vertices is a subgraph (therefore a minor)
of $\K_m$, which is a minor of $\K_{m,m}$, which is a subgraph
(therefore a minor) of $\K_{1,m,m}$.
This means that weighted maximum cut is still $\NP$-complete
if we restrict $G$ to be in the form of $\K_{1,m,m}$.

\subsection{Turing reduction from weighted maximum cut to membership test of
            cut polytopes}

Avis and Deza~\cite{AviDez-Net91} show the membership test to cut polytopes
of complete graphs is $\NP$-hard.
This implies the $\NP$-hardness of the membership test
to correlation polytopes of complete graphs.

In a similar way, we can prove the following theorem.

\begin{theorem} \label{thrm:cut-hard}
  The membership test to $\CutP(\K_{1,m,m})$ is $\NP$-hard.
\end{theorem}

\begin{proof}
  Let $\K_{1,m,m}=(V,E)$ and $P=\CutP(\K_{1,m,m})$.
  The optimization of a linear function on $P$ is
  equivalent to the weighted maximum cut problem on $\K_{1,m,m}$,
  so it is $\NP$-complete.
  We know $P$ is included in a hypercube $[0,1]^E$
  and $P$ includes a hypercube
  $[\frac{1}{\binom{2m+1}{2}+1},\frac{1}{\binom{2m+1}{2}}]^E$.
  The proof is completed by the polynomial-time Turing reduction
  from optimization problem to membership problem
  given by Corollary~4.3.12 and Theorem~6.3.2 (a) in \cite{GroLovSch:geo88}.
\end{proof}

Now the proof of Theorem~\ref{thrm:bell-hard} is obtained
by combining Theorems~\ref{thrm:bell-cut} and \ref{thrm:cut-hard}.

\section{Projections of Bell polytope and lifting their faces back}
  \label{sect:proj}

Let $\varphi\colon V\to U$ be an affine mapping
between two affine spaces $U$ and $V$.
If two polytopes $P\subseteq U$ and $Q\subseteq V$
satisfy $\varphi(Q)=P$,
an inequality
\begin{equation}
  \vct{a}^\trans\vct{u}\le a_0 \label{eq:lifting-1}
\end{equation}
is valid for $P$ if and only if the inequality
\begin{equation}
  \vct{a}^\trans\varphi(\vct{v})\le a_0 \label{eq:lifting-2}
\end{equation}
is valid for $Q$.
When \eqref{eq:lifting-1} is valid for $P$,
we call the valid inequality \eqref{eq:lifting-2} for $Q$
the \emph{lifting} of \eqref{eq:lifting-1} by the affine mapping $\varphi$.

For \eqref{eq:lifting-2} to support a facet of $Q$,
it is necessary for \eqref{eq:lifting-1} to support a facet of $P$.
Whether it is sufficient or not depends on $P$, $Q$ and $\varphi$.
If it is, we can obtain some of the facets of $Q$
by lifting the facets of the polytope $P$ with lower dimension,
which are hopefully computed easily.
Note that we cannot obtain all the facets of $Q$ this way;
if $Q$ has at least one vertex%
\footnote{We exclude the trivial case where
  the restriction of $\varphi$ to $Q$ is injective.},
$Q$ has at least one facet which is not in the form of \eqref{eq:lifting-2}.

In this section, we consider lifting by several specific projections
for the case where $Q$ is a Bell polytope.

\subsection{Projection from $\BellP(n+1,m,v)$
            to $\BellP(n,m,v)$}

Fix any $j=1,\dots,m$.
Let $\varphi\colon\R^{(mv)^{n+1}}\to\R^{(mv)^n}$
be a linear mapping which maps every point $\vct{q}'\in\R^{(mv)^{n+1}}$
to a point $\vct{q}\in\R^{(mv)^n}$ defined by
\[
  q_{(j_1,k_1),\dots,(j_n,k_n)}
  = \sum_{k=1}^v q'_{(j_1,k_1),\dots,(j_n,k_n),(j,k)}.
\]
Note that
if $\vct{q}'\in\BellP(n+1,m,v)$, $\varphi(\vct{q}')$ does not depend on
the choice of $j$%
\footnote{This is because any correlation table in $\BellP(n+1,m,v)$
  satisfies the no-signaling condition.}.
By the definition of $\BellP(n,m,v)$,
$\varphi(\BellP(n+1,m,v))=\BellP(n,m,v)$.

This means that if an inequality
\begin{equation}
  \vct{a}^\trans\vct{q}\le a_0 \label{eq:bell-lifting-n-1}
\end{equation}
is valid for $\BellP(n,m,v)$, the inequality
\begin{equation}
  \vct{a}^\trans\varphi(\vct{q})\le a_0 \label{eq:bell-lifting-n-2}
\end{equation}
obtained as the lifting of \eqref{eq:bell-lifting-n-1} by $\varphi$
is valid for $\BellP(n+1,m,v)$.
For example, let us consider the CHSH inequality~\cite{ClaHorShiHol-PRL69}:
\[
  q_{(1,1),(1,1)}-q_{(1,1),(2,2)}-q_{(2,2),(2,1)}-q_{(2,1),(1,1)}\le0,
\]
which supports a facet of $\BellP(2,2,2)$.
The lifting of the CHSH inequality by $\varphi$ is the inequality
\begin{multline}
  (q_{(1,1),(1,1),(1,1)}+q_{(1,1),(1,1),(1,2)})
  -(q_{(1,1),(2,2),(1,1)}+q_{(1,1),(2,2),(1,2)}) \\
  -(q_{(2,2),(2,1),(1,1)}+q_{(2,2),(2,1),(1,2)})
  -(q_{(2,1),(1,1),(1,1)}+q_{(2,1),(1,1),(1,2)})\le0.
  \label{eq:bell-lifting-n-3}
\end{multline}
By the fact described above, the inequality \eqref{eq:bell-lifting-n-3}
is valid for $\BellP(3,2,2)$.
However, the inequality \eqref{eq:bell-lifting-n-3}
does not support a facet of $\BellP(3,2,2)$ by the following theorem.

\begin{theorem} \label{thrm:bell-lifting-n}
  The inequality \eqref{eq:bell-lifting-n-2}
  never supports a facet of $\BellP(n+1,m,v)$.
\end{theorem}

\begin{proof}[Outline of proof]
  By normalization condition~\eqref{eq:normalization-condition},
  we can assume $a_0=0$ without loss of generality.

  Let $F$ be the face of $\BellP(n,m,v)$ supported by
  \eqref{eq:bell-lifting-n-1},
  and $F'$ be the face of $\BellP(n+1,m,v)$ supported by
  \eqref{eq:bell-lifting-n-2}.
  For $\vct{c}\in\{1,\dots,v\}^m$,
  define $\vct{a}^{\vct{c}}\in\R^{(mv)^{n+1}}$ by
  \[
    a^{\vct{c}}_{(j_1,k_1),\dots,(j_n,k_n),(j_{n+1},k_{n+1})}=\begin{cases}
      a_{(j_1,k_1),\dots,(j_n,k_n)} & \text{if $k_{n+1}=c_{j_{n+1}}$,} \\
      0                             & \text{otherwise.}
    \end{cases}
  \]
  Then $(\vct{a}^{\vct{c}})^\trans\vct{q}'=0$ for any vertex
  $\vct{q}'$ of $F'$ and any $\vct{c}\in\{1,\dots,v\}^m$.
  This means that the face $F'$ of $\BellP(n+1,m,v)$ lies in the
  intersection of the hyperplanes $\vct{a}^{\vct{c}}\vct{q}'=0$,
  and therefore $F'$ is not a facet of $\BellP(n+1,m,v)$.
\end{proof}

For example, the CHSH inequality is not a facet-supporting inequality for
$\BellP(n,2,2)$ for any $n>2$.

\subsection{Projection from $\BellP(n,m+1,v)$
            to $\BellP(n,m,v)$} \label{sbsect:proj-lifting-m}

\subsubsection{Symmetric setting}
By using a linear mapping $\varphi\colon\R^{((m+1)v)^n}\to\R^{(mv)^n}$
which maps every point $\vct{q}'\in\R^{((m+1)v)^n}$ to
a point $\vct{q}\in\R^{(mv)^n}$ defined by
\[
  q_{(j_1,k_1),\dots,(j_n,k_n)}
  = q'_{(k_1,j_1),\dots,(k_n,j_n)},
\]
we have $\varphi(\BellP(n,m+1,v))=\BellP(n,m,v)$.
Therefore, if an inequality
\begin{equation}
  \vct{a}^\trans\vct{q}\le a_0 \label{eq:bell-lifting-m-1}
\end{equation}
is valid for $\BellP(n,m,v)$, its lifting by $\varphi$
\begin{equation}
  \vct{a}^\trans\varphi(\vct{q})\le a_0 \label{eq:bell-lifting-m-2}
\end{equation}
is valid for $\BellP(n,m+1,v)$.
In contrast to the previous case of lifting to $\BellP(n+1,m,v)$,
this time the following theorem holds.

\begin{theorem} \label{thrm:bell-lifting-m}
  In case of $n=v=2$, if the inequality \eqref{eq:bell-lifting-m-1}
  supports a facet of $\BellP(n,m,v)$,
  its lifting \eqref{eq:bell-lifting-m-2} supports a facet
  of $\BellP(n,m+1,v)$.
\end{theorem}

\begin{proof}[Outline of proof]
  The proof is immediate from the affine isomorphism of Bell polytopes
  and cut polytopes:
  \begin{align*}
    \BellP(2,m,2)&\cong_{\aff}\CutP(\K_{1,m,m}), \\
    \BellP(2,m+1,2)&\cong_{\aff}\CutP(\K_{1,m+1,m+1}),
  \end{align*}
  and the 0-lifting theorem of cut polytope of graphs~\cite{Des-ORL90}.
\end{proof}

For example, the CHSH inequality supports a facet of $\BellP(2,m,2)$ for any
$m\ge2$.%
\footnote{As one direction of research,
  we can study the characteristics of facets of $\BellP(2,m+1,2)$
  which do not appear in $\BellP(2,m,2)$.}

\subsubsection{Asymmetric setting}\label{sec:asymm}
Both {\'{S}}liwa~\cite{Sli-PLA03} and Collins and
Gisin~\cite{ColGis-JPA04} consider the setting with asymmetric numbers
of observables independently. For these settings, Alice has $m_A$
observables and Bob has $m_B$. From similar argument to
Theorem~\ref{thrm:bell-lifting-m}, we can conclude that the tight Bell
inequality for $m_{A}, m_{B}$ is also tight for $ m_{A}^{\prime} \geq
m_{A},m_{B}^{\prime} \geq m_{B}$. Therefore, Bell inequalities
$I_{3422}^{k}, k=1,2,3$ of~\eqref{eq:I3422-2} and
\eqref{eq:I3422-13}~\cite{ColGis-JPA04} also
support facets of $\BellP(2,m,2), m \geq 4$.

\subsection{Projection from $\BellP(n,m,v+1)$
            to $\BellP(n,m,v)$}

We define a mapping $\psi\colon\{1,\dots,v+1\}\to\{1,\dots,v\}$ by
\[
  \psi(k')=\begin{cases}
    k' & \text{if $1\le k'\le v-1$,} \\
    v  & \text{if $k'\in\{v,v+1\}$.}
  \end{cases}
\]
By using a linear mapping $\varphi\colon\R^{(m(v+1))^n}\to\R^{(mv)^n}$
which maps every point $\vct{q}'\in\R^{(m(v+1))^n}$ to the point
$\vct{q}\in\R^{(mv)^n}$ defined by
\[
  q_{(j_1,k_1),\dots,(j_n,k_n)}
  = \sum_{\substack{\vct{k}'\in\{1,\dots,v+1\}^n \\
                    \psi(k'_i)=k_i\;(1\le\forall i\le n)}}
      q'_{(j_1,k'_1),\dots,(j_n,k'_n)},
\]
we see that $\varphi(\BellP(n,m,v+1))=\BellP(n,m,v)$.

We have not yet determined whether this operation is facet-preserving.

\subsection{Projection from correlation tables
            to correlation functions}
  \label{sbsect:proj-correlation-function}

For every $\vct{q}\in\R^{(mv)^n}$, consider a point $\vct{s}\in\R^{m^n v}$
defined by
\[
  s_{j_1\dotsm j_n k}
  =\sum_{\substack{\vct{k}\in\{1,\dots,v\}^n \\ k_1+\dots+k_n\equiv k\pmod{v}}}
    q_{(j_1,k_1),\dots,(j_n,k_n)},
\]
and denote this point $\vct{s}$ by $\varphi(\vct{q})$.
This defines a linear mapping $\varphi\colon\R^{(mv)^n}\to\R^{m^n v}$.

For every correlation table $\vct{q}\in\BellP(n,m,v)$ induced
by a classical $(n,m,v)$-system,
the point $\varphi(\vct{q})$ is called
the \emph{full correlation function}
(in \cite{WerWol-PRA01} in $(n,2,2)$ case)
or \emph{correlation function}
(in \cite{ColGisLinMasPop-PRL02} in $(2,2,v)$ case)
defined by $\vct{q}$.

Let $\calW(n,m,v)=\varphi(\BellP(n,m,v))$.
Werner and Wolf~\cite{WerWol-PRA01} show
that $\calW(n,2,2)$ is affinely isomorphic
to the $2^n$-dimensional crosspolytope.
In $(2,2,2)$ case, the facets of the crosspolytope correspond to
trivial and CHSH inequalities of $\BellP(2,2,2)$, and the facets of
the crosspolytope Werner and Wolf consider can be seen as
generalization of these inequalities.
By Theorem~\ref{thrm:bell-lifting-n}, none of them are lifted to
facets of $\BellP(n,2,2)$.
Collins, Gisin, Linden, Massar and Popescu~\cite{ColGisLinMasPop-PRL02}
give a valid inequality for $\calW(2,2,v)$,
which is later called the CGLMP inequality,
and Masanes~\cite{Mas-QIC03} shows that the lifting of the CGLMP inequality
by $\varphi$ is a facet of $\BellP(2,2,v)$.
It is not known whether the lifting of a facet of $\calW(n,m,v)$
always supports a facet of $\BellP(n,m,v)$.

\section*{Acknowledgment}
We thank N.~Gisin for helpful comments on the original version of this
paper.


\begin{thebibliography}{22}
\expandafter\ifx\csname natexlab\endcsname\relax\def\natexlab#1{#1}\fi
\expandafter\ifx\csname url\endcsname\relax
  \def\url#1{{\tt #1}}\fi

\bibitem[Avis()]{Avi:lrs}
D.~Avis.
\newblock lrs.
\newblock URL \url{http://cgm.cs.mcgill.ca/~avis/C/lrs.html}.

\bibitem[Avis and Deza(1991)]{AviDez-Net91}
D.~Avis and M.~Deza.
\newblock The cut cone, {$L^1$} embeddability, complexity and multicommodity
  flows.
\newblock {\em Networks}, 21:\penalty0 595--617, 1991.

\bibitem[Clauser et~al.(1969)Clauser, Horne, Shimony, and
  Holt]{ClaHorShiHol-PRL69}
J.~F. Clauser, M.~A. Horne, A.~Shimony, and R.~A. Holt.
\newblock Proposed experiment to test local hidden-variable theories.
\newblock {\em Physical Review Letters}, 23\penalty0 (15):\penalty0 880--884,
  Oct. 1969.

\bibitem[Collins and Gisin(2004)]{ColGis-JPA04}
D.~Collins and N.~Gisin.
\newblock A relevant two qubit {B}ell inequality inequivalent to the {CHSH}
  inequality.
\newblock {\em Journal of Physics A: Mathematical and General}, 37\penalty0
  (5):\penalty0 1775--1787, Feb. 2004.

\bibitem[Collins et~al.(2002)Collins, Gisin, Linden, Massar, and
  Popescu]{ColGisLinMasPop-PRL02}
D.~Collins, N.~Gisin, N.~Linden, S.~Massar, and S.~Popescu.
\newblock {B}ell inequalities for arbitrarily high-dimensional systems.
\newblock {\em Physical Review Letters}, 88\penalty0 (040404), Jan. 2002.

\bibitem[{\uppercase{d}e}~Simone(1990)]{Des-ORL90}
C.~{\uppercase{d}e}~Simone.
\newblock Lifting facets of the cut polytope.
\newblock {\em Operations Research Letters}, 9\penalty0 (5):\penalty0 341--344,
  Sept. 1990.

\bibitem[{\uppercase{d}e}~Simone et~al.(1994){\uppercase{d}e}~Simone, Deza, and
  Laurent]{DesDezLau-DM94}
C.~{\uppercase{d}e}~Simone, M.~Deza, and M.~Laurent.
\newblock Collapsing and lifting for the cut cone.
\newblock {\em Discrete Mathematics}, 127\penalty0 (1--3):\penalty0 105--130,
  Mar. 1994.

\bibitem[Deza and Laurent(1997)]{DezLau:cut97}
M.~M. Deza and M.~Laurent.
\newblock {\em Geometry of Cuts and Metrics}, volume~15 of {\em Algorithms and
  Combinatorics}.
\newblock Springer, May 1997.

\bibitem[Fine(1982)]{Fin-PRL82}
A.~Fine.
\newblock Hidden variables, joint probability, and the {B}ell inequalities.
\newblock {\em Physical Review Letters}, 48\penalty0 (5):\penalty0 291--295,
  Feb. 1982.

\bibitem[Garey and Johnson(1979)]{GarJoh:computers79}
M.~R. Garey and D.~S. Johnson.
\newblock {\em Computers and Intractability: A Guide to the Theory of
  {NP}-Completeness}.
\newblock W.~H.~Freeman, June 1979.
\newblock ISBN 0-7167-1045-5.

\bibitem[Grishukhin(1990)]{Gri-EJC90}
V.~P. Grishukhin.
\newblock All facets of the cut cone {$C_n$} for {$n=7$} are known.
\newblock {\em European Journal of Combinatorics}, 11:\penalty0 115--117, 1990.

\bibitem[Gr{\"{o}}tschel et~al.(1988)Gr{\"{o}}tschel, Lov{\'{a}}sz, and
  Schrijver]{GroLovSch:geo88}
M.~Gr{\"{o}}tschel, L.~Lov{\'{a}}sz, and A.~Schrijver.
\newblock {\em Geometric Algorithms and Combinatorial Optimization}, volume~2
  of {\em Algorithms and Combinatorics}.
\newblock Springer, 1988.

\bibitem[Masanes(2003)]{Mas-QIC03}
{\relax{Ll}}.~Masanes.
\newblock Tight {B}ell inequality for $d$-outcome measurements correlations.
\newblock {\em Quantum Information \& Computation}, 3\penalty0 (4):\penalty0
  345--358, July 2003.

\bibitem[Peres(1999)]{Per:all99}
A.~Peres.
\newblock All the {B}ell inequalities.
\newblock {\em Foundations of Physics}, 29\penalty0 (4):\penalty0 589--614,
  Apr. 1999.

\bibitem[Pitowsky(1989)]{Pit:prob89}
I.~Pitowsky.
\newblock {\em Quantum Probability --- Quantum Logic}, volume 321 of {\em
  Lecture Notes in Physics}.
\newblock Springer, 1989.

\bibitem[Pitowsky(1991)]{Pit-MP91}
I.~Pitowsky.
\newblock Correlation polytopes: Their geometry and complexity.
\newblock {\em Mathmatical Programming}, 50:\penalty0 395--414, 1991.

\bibitem[Pitowsky and Svozil(2001)]{PitSvo-PRA01}
I.~Pitowsky and K.~Svozil.
\newblock Optimal tests of quantum nonlocality.
\newblock {\em Physical Review A}, 64\penalty0 (014102), June 2001.

\bibitem[{Research Group Discrete Optimization, University of
  Heidelberg}()]{Smapo}
{Research Group Discrete Optimization, University of Heidelberg}.
\newblock {SMAPO}---``small'' 0/1-polytopes in combinatorial optimization.
\newblock URL
  \url{http://www.iwr.uni-heidelberg.de/groups/comopt/software/SMAPO/cut/cut.html}.

\bibitem[{\'{S}}liwa(2003)]{Sli-PLA03}
C.~{\'{S}}liwa.
\newblock Symmetries of the {B}ell correlation inequalities.
\newblock {\em Physics Letters A}, 317\penalty0 (3--4):\penalty0 165--168, Oct.
  2003.

\bibitem[Werner and Wolf(2001{\natexlab{a}})]{WerWol-PRA01}
R.~F. Werner and M.~M. Wolf.
\newblock All-multipartite {B}ell-correlation inequalities for two dichotomic
  observables per site.
\newblock {\em Physical Review A}, 64\penalty0 (032112), Aug.
  2001{\natexlab{a}}.

\bibitem[Werner and Wolf(2001{\natexlab{b}})]{WerWol-QIC01}
R.~F. Werner and M.~M. Wolf.
\newblock Bell inequalities and entanglement.
\newblock {\em Quantum Information \& Computation}, 1\penalty0 (3):\penalty0
  1--25, Oct. 2001{\natexlab{b}}.

\bibitem[Ziegler(1998)]{Zie:lectures98}
G.~M. Ziegler.
\newblock {\em Lectures on Polytopes}, volume 152 of {\em Graduate Texts in
  Mathematics}.
\newblock Springer, revised edition, 1998.

\end{thebibliography}
\end{document}